\documentclass[aps,pre,twocolumn,showpacs]{revtex4-1}

\usepackage{bm}
\usepackage{graphicx}
\usepackage{amssymb,amsmath}
\usepackage{amsfonts}
\usepackage{amsthm} 
\usepackage{leftidx}
\usepackage{stmaryrd}  

\newcommand*{\be}{\begin{equation}}
\newcommand*{\ee}{\end{equation}}

\providecommand*{\abs}[1]{\lvert#1\rvert}

\newcommand{\sump}[2]{{\sum_{#1}}\raise4pt\hbox{\hskip#2$'$}}

\newcommand{\bb}{b}
\newcommand{\bs}[1]{\boldsymbol{#1}}

\begin{document}
\title{Path Integral Approach to  Random Neural Networks}

\date{V 2.2.2 2018/11/09 AC}


\author{A. Crisanti}
\email{andrea.crisanti@uniroma1.it}
\affiliation{Department of Physiscs, and Institute of Complex Systems (ISC-CNR), 
University ``La Sapienza'', 
                 P.le Aldo Moro 2, I-00185 Roma, Italy}
\author{H. Sompolinksy}
\affiliation{Racah Institute of Physics and The Edmond and Lily Safra Center for Brain Sciences,
The Hebrew University of Jerusalem, 9190401, Israel}

\begin{abstract}
In this work we study of the dynamics of large size random neural networks. Different methods have 
been developed to analyse their behavior, most of them rely on heuristic methods based on 
Gaussian assumptions regarding the fluctuations in the limit of infinite sizes.  These approaches, 
however, do not justify the underlying assumptions systematically. Furthermore, they are incapable 
of deriving in general the stability of the derived mean field equations, and they are not amenable to 
analysis of finite size corrections. Here we present a systematic method based on Path Integrals 
which overcomes these limitations. We apply the method to a large non-linear rate based neural 
network with random asymmetric connectivity matrix. We derive the Dynamic Mean Field (DMF) 
equations for the system, and derive the Lyapunov exponent of the system. Although the main 
results are well known, here for the first time, we calculate the spectrum of fluctuations around the 
mean field equations from which we derive the general stability conditions for the DMF states. The 
methods presented here, can be applied to neural networks with more complex dynamics and 
architectures. In addition, the theory can be used to compute systematic finite size corrections to 
the mean field equations. 
\end{abstract} 


\maketitle

\section{Introduction}
The present paper aims to present a detailed derivation of the Path Integral framework for the study 
of the dynamical properties of neural networks.  This framework can be applied to a broad spectrum 
of network models.  For concreteness, we shall consider a class of model, extending the model first 
introduced in 1972 by Amari \cite{A72}. This class of model is simple enough to allow for a full 
analytical description yet presenting a highly non-trivial dynamical behavior. In the models 
considered here, the state of the $i$-th neuron of the network at time $t$ is represented by a 
continuous “spin” variable $S_i(t)$ which represents the firing activity of the neuron. The state of a 
neuron is determined by the ``post-synaptic'' potential $h_i(t)$ acting on it through the relationship 
$S_i(t) = \phi(g h_i(t))$, where $g$ is a gain parameter measuring the gain of the response. The 
function $\phi(x)$ is usually a sigmoid function which defines the input/output characteristic of the 
neurons. As a concrete example we shall consider $\phi(x) = \tanh(x)$ as prototype of generic odd 
symmetric saturated sigmoid functions satisfying:  $\phi(x) = -\phi(-x)$, $\phi(\pm\infty) = \pm 1$, 
and $\phi’(0) = d\phi(x)/dx|_{x=0}=1$ so that $g$ is the slope of the linear response of the neuron to 
small post-synaptic potential.  The theory can be easily extended to transfer functions which are not 
odd symmetric. The case of non-saturated transfer functions have been studied recently 
\cite{KadSom15,Barraetal18}.

The dynamical behaviour of a network of $N$ neurons is governed by
the first order differential equations: 
\begin{equation}
\frac{d}{dt}\,h_{i}(t)=-h_{i}(t)+\sum_{j=1}^{N}J_{ij}\,S_{i}(t),\qquad i=1,\dotsc,N.\label{eq:kirch}
\end{equation}
In electrical terms Eqs. \eqref{eq:kirch} are the Kirchhoff current law
of the neuron, where the current charging the membrane capacitance,
the l.h.s term, must equal the current through the membrane resistance,
first term in the r.h.s, plus the current due to the activity of the
other cells, last term in the r.h.s. For simplicity the microscopic
time constant is taken equal to one.

The (real) matrix $J_{ij}$, with $J_{ii}=0$, gives the properties
of the synaptic coupling between the pre-synaptic $j$-th neurons
and the post-synaptic $i$-th neuron. It defines the topology of the
network: $J_{ij}=0$ not connected $J_{ij}\not=0$ connected; the
type of the synaptic connection: $J_{ij}>0$ excitatory $J_{ij}<0$
inhibitory; the strength of the connection: $\abs{J_{ij}}$.

We shall focus on the steady state of the network, that is the dynamical
state in which the network settles down after a reasonable time has
elapsed from the initial time $t_{0}$. Thus we shall assume that
$t_{0}\to-\infty$ so that memory of the initial state at $t_{0}$
has been lost.

Clearly the dynamical behaviour of the network depends on $J_{ij}$.
Nevertheless we can distinguish two classes. 
If the matrix $J_{ij}$ is \textit{symmetric}, i.e., $J_{ji} = J_{ij}$, then the dynamical 
equations \eqref{eq:kirch} describes the relaxation 
\begin{equation}
  \frac{d}{dt} h_i(t) = -\frac{\partial}{\partial h_i}E(h_1,\dotsc,h_N) \Bigl|_{h_i = h_i(t)}
\end{equation}
of the energy function:
\begin{equation}
  E(h_1,\dotsc,h_N) = \frac{1}{2}\sum_i h_i^2 - \frac{1}{2}\sum_{ij} J_{ij}\, S_i S_i.
\end{equation}
The dynamics hence converges towards stable fixed points which correspond
to the stable local minima of $E(h_1,\dotsc, h_N)$. The structure of the fixed points can be 
complex \cite{Hop,SK},
nevertheless the asymptotically long time state is simple.

 If the matrix is \textit{non-symmetric},
i.e., $J_{ji}\not=J_{ij}$, an energy function cannot be defined and
a richer steady state behaviour emerges: besides fix points, limit
cycles and chaotic behaviour are also possible.

We shall consider here the simple case of a fully connected network
with \textit{random}, \textit{asymmetric} and \textit{independent}
couplings: 
\begin{equation}
\overline{J_{ij}}=0,
\quad
\overline{(J_{ij})^{2}}=1/N,
\quad
\overline{J_{ij}\,J_{ji}}=0,
\quad i\not=j.
\label{eq:coup_mom}
\end{equation}
Here, and in the following, $\overline{(\dotso)}$ denotes averaging with the coupling probability
distribution $P(\mathbf{J})=\prod_{ij}P(J_{ij})$. The scaling of
the second moment with $N$ ensures that the second term on the r.h.s.
in \eqref{eq:kirch} is $O(1)$ as $N\to\infty$ (thermodynamic limit).

The assumption of zero average implies that there is
not a preferred type of synaptic connection. This can be relaxed by
imposing a finite average $J_{0}/N$ to tune preferred inhibitory
$(J_{0}<0)$ or excitatory $(J_{0}>0)$ synaptic connections.

Provided the high order moments of $P(\mathbf{J})$ do not grow too fast
with $N$, in the large $N$ limit only the first two moments are
needed. Thus we can assume that $J_{ij}$ are i.i.d Gaussian variables.

The full solution of the model, referred to as Dynamic Mean Field Theory (DMFT),  
has been presented and discussed in Ref. \cite{SomCriSom88}. Since then, several variations of this model has been
studied,  see e.g. Refs.
\cite{MolSchSch92,CheAma01,SamCes07,ToyRahPan09,RajAbbSom10,AljSteSha15,KadSom15,%
SteSomAbb14,BraSolOpp16}.
This model has served also as  the basis for computational modeling in
recurrent networks in particular work on Echo State networks, reservoir
computing, Force Learning.

Although the DMFT can be derived by an intuitive construction of self-consistent
equations for the fluctuations in the system, these ad-hoc derivation
suffers from potentially severe limitations. Most importantly, determining
the stability conditions for the network dynamical state is a considerable
challenge for such a naive approach. Also, computing various response
and correlation functions require going beyond the DMFT themselves.
Finally, extensions to more complex architecture or dynamics may be
less amenable to naive approaches to the construction of the correct
self consistent DMFT equations. 
Least but not last, it is hard to compute corrections
to the theory without a more systematic formalism. 
Here we present
a systematic approach to the study of dynamical states in random neural
networks using Path Integral Method. Path integrals have been
extensively used in the study of stochastic dynamics in statistical
mechanics, from the pioneering work of the Martin-Siggia-Rose
\cite{MarSigRos73} to work
on critical phenomena and RG analysis \cite{DeDom76,Jan76,DeDomPel78}
and to study the stochastic dynamics of spin glasses
\cite{DeDom78,SomZip82,CriSom87}.

The study of deterministic dynamical systems with Path Integrals, such as in the
present study is less common. Nevertheless, in our case this application
is facilitated by the presence of asynchronous chaotic state which
generates dynamical deterministic fluctuations with stationary statistics.
The present approach, which was used to derive the results reported in Ref.
\cite{SomCriSom88},  
expands on unpublished manuscript by the same authors from 1988.
For a related approach see Ref. \cite{Schu16}.
For an alternative study of neural networks based on the analogy with
conservative Newtonian dynamics see, e.g., Ref. \cite{BarBecFac18}. 

The plan of the paper is as follows. In Section \ref{sec:DFT_G} we derive
the DFT describing the dynamical behaviour of the model \eqref{eq:kirch}.
The possible different solutions of the DMFT valid in the limit $N\gg 1$ are discussed in Section 
\ref{sec:DMFT_Sol},
and  their stability analysed in Section \ref{sec:DMFT_Stab}. 
Finally in Section  \ref{sec:Lyap}, as an illustration of how dynamical quantities can be 
computed using DFT,  
we present the calculation of the maximum Lyapunov exponent.

\section{\label{sec:DFT_G}Dynamical Field Theory (DFT)}
In this Section we shall show how the the dynamical
behaviour of the network can be described using path integral methods.
Prior to this we introduce the useful shorthand notation $h_{i}^{a}=h_{i}(t_{a})$
and rewrite the equation of motion \eqref{eq:kirch}  as: 
\begin{equation}
\partial_{a}h_{i}^{a}=-h_{i}^{a}+\sum_{j=1}^{N}\,J_{ij}\,S_{j}^{a},\label{eq:eq_dyn}
\end{equation}
where $\partial_{a}=(d/dt_{a})+\delta$ ($\delta\to0^{+}$) to ensure
causality \cite{Note:Causality}, and $S_{i}^{a}=\phi(gh_{i}^{a})$.

\subsection{Path Integral and Dynamical Field Theory}
The strategy of the path integral approach is to
derive a generating functional for the relevant correlation and response
functions induced by the dynamics \eqref{eq:eq_dyn}.
To work on a finite dimensional space, one starts by dividing the
time interval of interest $[t_{0},t]$ into $n$ segments of length
$\delta t$ and changing the differential equation $\partial_{a}h_{i}^{a}=f(h_{i}^{a})$
into the finite-difference equation: 
\begin{equation}
    h_{i}^{a+1}-h_{i}^{a}=f(h_{i}^{a})\,\delta t+\bb_{i}^{a}\delta t+h_{i}^{0}\delta^{\rm Kr}_{a0},
\label{eq:eq_dyn_dsc}
\end{equation}
with the (discrete) index $a=0,1,\dotsc,n$ indicating the time. Two
terms have been added: an external field $\bb_{i}^{a}$ to evaluate
response functions and the initial condition $h_{i}^{0}\delta_{a0}$, where $\delta^{\rm Kr}_{ab}$
is the Kronecker delta, to enforce the initial condition at $t_{0}$.
The continuum limit is recovered by taking $n\to\infty$ and $\delta t\to0$
with $n\delta t$ fixed \cite{Note:Delta}.

Denoting by $\tilde{h}_{i}^{a}$ the solution to Eq. \eqref{eq:eq_dyn_dsc},
the generating functional of correlation and response functions of the
dynamical system \eqref{eq:eq_dyn_dsc} reads: 
\begin{widetext}
\begin{equation}
  \begin{split}
       Z[\hat{\bb},\bb] & =\int\prod_{a}\prod_{i}dh_{i}^{a}\,
           \delta\bigr(h_{i}^{a}-\tilde{h}_{i}^{a}\bigl)\,e^{-i\hat{\bb}_{i}^{a}h_{i}^{a}}
           \\
                 & =\int\prod_{a}\prod_{i}dh_{i}^{a}\,
                \delta\bigr(h_{i}^{a+1}-h_{i}^{a}-f(h_{i}^{a})\delta t-\bb_{i}^{a}\delta t-h_{i}^{0}\delta_{a0}\bigl)
                \,e^{-i\hat{\bb}_{i}^{a}h_{i}^{a}}.
\end{split}
\label{eq:delta_id}
\end{equation}
\end{widetext}
The second line is obtained by using the identity 
$\delta\bigl(h_{i}^{a}-\tilde{h}_{i}^{a}\bigr)=|F'(h_{i}^{a})|\ \delta\bigl[F(h_{i}^{a})\bigr]$
where 
$F(h_{i}^{a})=h_{i}^{a+1}-h_{i}^{a}-f(h_{i}^{a})\delta t-\bb_{i}^{a}\delta t-h_{i}^{0}\delta^{\rm Kr}_{a0}$
and $F'(h_{i}^{a})$ is the Jacobian of the transformation $h_{i}^{a}-\tilde{h}_{i}^{a}=0\to F(h_{i}^{a})=0$.
The Jacobian depends on the discretisation scheme used to translate
the differential equation into a finite-difference equation, even
in the continuum limit $n\to\infty$ \cite{HuntRoss81}. The role
of the Jacobian is to ensure that correlation and response functions
do not depend on the discretisation scheme used to construct the generating
functional, apart form the initial value of the response functions
\cite{Honkonen11}. The scheme adopted in Eq. \eqref{eq:eq_dyn_dsc},
known as the Ito scheme in the theory of Stochastic Differential Equations,
has the advantage that the Jacobian is equal to one. Another consequence
of this scheme is that $\theta(0^{-})=0$ and $\theta(0^{+})=1$,
where $\theta(x)$ is the Heaviside step function. See, e.g., Ref. \cite{Gardiner85}.

The expression \eqref{eq:delta_id} can be made more manageable by
using the Fourier representation of the Dirac $\delta$-function 
\begin{equation}
\delta(z_{i}^{a})=\int_{-\infty}^{+\infty}\frac{d\hat{h}_{i}^{a}}{2\pi}\,e^{-i\hat{h}_{i}^{a}\,z_{i}^{a}},
\end{equation}
to rewrite it as: 
\begin{equation}
\begin{split}Z[\hat{\bb},\bb]=&\int  \prod_{a}\prod_{i}\frac{d\hat{h}_{i}^{a}\,dh_{i}^{a}}{2\pi}\,\\
 & \times
 e^{-i\hat{h}_{i}^{a}\,\bigl(h_{i}^{a+1}-h_{i}^{a}-f(h_{i}^{a})\delta t-\bb_{i}^{a}\delta t-h_{i}
 ^{0}\delta_{a0}\bigr) +i\hat{\bb}_{i}^{a}h_{i}^{a}}.
\end{split}
\end{equation}
Taking the continuum limit $n\to\infty$ with $n\delta t=t-t_{0}$,
$Z[\hat{\bb},\bb]$ becomes a path integral over all possible paths
$\{\hat{h}_{i},h_{i}\}_{t_{a}\in[t_{0},t]}$: 
\begin{equation}
\begin{split}Z[\hat{\bb},\bb]=\int\prod_{i}{\cal D}\hat{h}_{i}\,{\cal D}h_{i}\,
e^{-S[\hat{h},h]+\sum_{ia}\bigl(i\hat{\bb}_{i}^{a}h_{i}^{a}+i\hat{h}_{i}^{a}\bb_{i}^{a}\bigr)},
\end{split}
\label{eq:gen_fun}
\end{equation}
where 
${\cal D}\hat{h}_{i}=\lim_{n\to\infty}\prod_{a}d\hat{h}_{i}^{a}/2\pi$,
${\cal D}h_{i}=\lim_{n\to\infty}\prod_{a}dh_{i}^{a}$,
$\sum_{a}\equiv\int dt_{a}$ and $S[\hat{h},h]$ is the dynamical
action: 
\begin{equation}
\begin{split}S[\hat{h},h] & =\sum_{ia}i\hat{h}_{i}^{a}\,\bigl(\partial_{a}h_{i}^{a}-f(h_{i}^{a})-h_{i}^{0}\delta_{a0}\bigr)\\
 & =\sum_{ia}i\hat{h}_{i}^{a}\,\bigl(\partial_{a}h_{i}^{a}+h_{i}^{a}-\sum_{j}J_{ij}S_{j}^{a}-h_{i}^{0}\delta_{a0}\bigr),
\end{split}
\label{eq:msr_act}
\end{equation}
of the equation of motion \eqref{eq:eq_dyn} with initial condition
$h_{i}^{0}=h_{i}(t_{0})$. Here $\delta_{a0}\equiv\delta(t_{a}-t_{0})$.
We have not included the term $\bb_{i}^{a}$ into the dynamical
action because the original problem does not have an external field.


From the definition \eqref{eq:delta_id} it follows that $Z[0,\bb]=1$,
then 
\begin{equation}
\begin{split}
       \frac{\delta}{\delta i\hat{\bb}_{i_{1}}^{a_{1}}}\dotsc \frac{\delta}{\delta i\hat{\bb}_{i_{n}}^{a_{n}}}
       \frac{\delta}{\delta\bb_{j_{1}}^{b_{1}}}\dotsc & \frac{\delta}{\delta\bb_{j_{m}}^{b_{m}}}
        Z[\hat{\bb},\bb]\Bigr|_{\hat{\bb}=\bb=0}
       \\
       & 
       =\bigl\langle h_{i_{1}}^{a_{1}}\dotsc h_{i_{n}}^{a_{n}}\hat{h}_{j_{1}}^{b_{1}}\dotsc\hat{h}_{j_{m}}^{b_{m}}\bigr\rangle_{J},
       \end{split}
\label{eq:corr_fun}
\end{equation}
are the correlation functions of $h_{i}^{a}$ and $\hat{h}_{i}^{a}$
over the dynamics generated by the action \eqref{eq:msr_act} for
fixed coupligs $J_{ij}$. 

Correlations of only $h$-fields are correlation functions of the
dynamics \eqref{eq:eq_dyn}. Those involving both $\hat{h}$ and $h$ fields the response
functions, as can be inferred from Eq. \eqref{eq:gen_fun} by noticing
that 
\begin{equation}
\bigl\langle h_{i_{1}}^{a_{1}}\dotsc h_{i_{n}}^{a_{n}}\hat{h}_{j_{1}}^{b_{1}}\dotsc\hat{h}_{j_{m}}^{b_{m}}\bigr\rangle_{J}=\frac{\delta}{\delta\bb_{j_{1}}^{b_{1}}}\dotsc\frac{\delta}{\delta\bb_{j_{m}}^{b_{m}}}\langle h_{i_{1}}^{a_{1}}\dotsc h_{i_{n}}^{a_{n}}\rangle_{J\bb}\Bigr|_{\bb=0},
\end{equation}
where the average $\langle(\dotso)\rangle_{J\bb}$ is over all
trajectory generated by the equation of motion \eqref{eq:eq_dyn}
in presence of the external field $\bb_{i}^{a}$. For this reason
hat-fields are also called \textit{response}-fields.
Note that since 
$(\delta/\delta b) Z[\hat{\bb}, \bb]|_{\hat{\bb}=\bb =0} = (\delta/\delta b) Z[0, \bb]|_{\bb =0}$
and $Z[0,\bb]=1$ 
correlations
involving only $\hat{h}$-fields vanish.

The correlation functions \eqref{eq:corr_fun} depend on the coupling
matrix $J_{ij}$ and are random quantities. Since $Z[0,0]=1$ 
averaged correlation functions can be obtained by averaging $Z[\hat{\bb},\bb]$
over the couplings $J_{ij}$ \cite{DeDom78}. 
In the case of the i.i.d Gaussian $J_{ij}$ \eqref{eq:coup_mom} this leads to 
\begin{equation}
\begin{split}
\overline{Z}[\hat{\bb},\bb] 
          = &\int\prod_{i}{\cal D}\hat{h}_{i}\,{\cal D}h_{i}\,
              \exp\Bigl[-\sum_{ia}  i\hat{h}_{i}^{a}\,(1+\partial_{a})h_{i}^{a}
              \\
              &+\frac{1}{2N}\sum_{ij}\Bigl(\sum_{a}i\hat{h}_{i}^{a}S_{j}^{a}\Bigr)^{2}
 -h_{i}^{0}\delta_{a0}
 \\
 & +\sum_{ia}\bigl(i\hat{\bb}_{i}^{a}h_{i}^{a}+i\hat{h}_{i}^{a}\bb_{i}^{a}\bigr)\Bigr].
\end{split}
\label{eq:Z_av}
\end{equation}
The non-local term can be simplified by introducing $C^{ab}=\sum_{i}S_{i}^{a}S_{i}^{b}/N$
\cite{SomZip82}
using the identity \cite{Note1}:
\begin{equation}
\begin{split}1 & =\int dC^{ab}\,\delta\Bigl(C^{ab}-\sum_{i}S_{i}^{a}S_{i}^{b}/N\Bigr)\\
 & =\int\frac{N}{2\pi}d\hat{C}^{ab}\,dC^{ab}\,\exp\left[-\frac{1}{2}i\hat{C}^{ab}\Bigl(NC^{ab}-\sum_{i}S_{i}^{a}S_{i}^{b}\Bigr)\right],\quad.
\end{split}
\end{equation}
The exponent in Eq.  \eqref{eq:Z_av} becomes diagonal in the site
index $i$ with a residual
site dependence due to the auxiliary fields $\bb_{i}^{a}$ and $\hat{\bb}_{i}^{a}$, 
because the system is fully connected.
The averaged generating functional can then be written as the partition
function 
\begin{equation}
\overline{Z}[\hat{\bb},\bb]=\int{\cal D}\hat{C}\,{\cal D}C\,e^{-N{\cal L}[\hat{C},C;\hat{\bb},\bb]},\label{eq:gen_fun_ave}
\end{equation}
of a dynamical field theory for the fields $\{\hat{C}^{ab},C^{ab}\}$,
with $\hat{C}^{ba}=\hat{C}^{ab}$ and $C^{ba}=C^{ab}$, described
by the action 
\begin{equation}
{\cal L}[\hat{C},C;\hat{\bb},\bb]=\frac{1}{2}\sum_{ab}\bigl(i\hat{C}^{ab}C^{ab}
          +\frac{\epsilon}{2}\,\hat{C}^{ab}\hat{C}^{ab}\bigr)-W[\hat{C},C;\hat{\bb},\bb],\label{eq:DFT_S}
\end{equation}
where 
\begin{equation}
\begin{split}
NW[\hat{C},C;\hat{\bb},\bb]=\sum_{i}\ln&\int {\cal D}\hat{h}_{i}\,{\cal D}h_{i}
    \\
    & \times
    e^{-S[\hat{h}_{i},h_{i};C,\hat{C}]+\sum_{a}(i\hat{h}_{i}^{a}\bb_{i}^{a}+i\hat{\bb}_{i}^{a}h_{i}^{a})},
\end{split}
    \label{eq:DFT_W}
\end{equation}
and 
\begin{equation}
\begin{split}S[\hat{h}_{i},h_{i};\hat{C},C]=\sum_{a}&\Bigl[i\hat{h}_{i}^{a}(1+  \partial_{a})h_{i}^{a}-h_{i}^{0}\delta_{a0}\Bigr]\\
 & -\frac{1}{2}\sum_{ab}\Bigl[i\hat{C}^{ab}S_{i}^{a}S_{i}^{b}+C^{ab}i\hat{h}_{i}^{a}i\hat{h}_{i}^{b}\Bigr].
\end{split}
\label{eq:DFT_hh_S}
\end{equation}
We have added a small regularising term $\epsilon\to0^{+}$ in Eq. \eqref{eq:DFT_S}
to make integrals well definite \cite{Note2}.
Note that the functional $W[\hat{C},C;\hat{\bb},\bb]$ is the generating
functional of connected (time) correlation functions of $h_{i}$ and $\hat{h}_{i}$
generated by the action: 
\begin{equation}
       {\cal L}[\hat{h},h;\hat{C},C,\hat{\bb},\bb]=
          \sum_{i}S[\hat{h}_{i},h_{i};C,\hat{C}]
         -\sum_{ia}(i\hat{h}_{i}^{a}\bb_{i}^{a}+i\hat{\bb}_{i}^{a}h_{i}^{a}).
\end{equation}
The relevant time correlation and response functions of the field
$S^{a}$ along the dynamical evolution governed by Eq. \eqref{eq:eq_dyn}
can be obtained from averages of $\hat{C}^{ab}$ and $C^{ab}$ with
the action ${\cal L}[\hat{C},C;0,0]$. For example, 
\begin{equation}
    \frac{1}{N}\sum_{i=1}^{N}\overline{\langle S_{i}(t_{a})S_{i}(t_{b})\rangle}_{J}
        =\langle S^{a}S^{b}\rangle=\langle C^{ab}\rangle,
\end{equation}
where $\langle(\dotso)\rangle$ denotes DFT average with action  ${\cal L}[\hat{C},C;0,0]$. 
Details are
in Appendix \ref{app:eq_C_ab}. The last equality follows because if $\hat{\bb}=\bb=0$,
or more generally if they do not depend on the site, different sites
decouples and are all equivalent.

The results obtained so far are valid for any value of $N$. In the
rest of this paper we shall consider the (thermodynamic) limit $N\gg1$.

\subsection{\label{sec:DMFT}Thermodynamic Limit and Dynamical Mean Field Theory.}
In the limit $N\to\infty$ the integral in Eq. \eqref{eq:gen_fun_ave}
is dominated by the largest value of the exponent. Therefore 
\begin{equation}
\overline{Z}[\hat{\bb},\bb]\sim\overline{Z}_{0}[\hat{\bb},\bb]=e^{-N{\cal L}_{0}[\hat{C},C;\hat{\bb},\bb]},\qquad N\gg1,
\end{equation}
where ${\cal L}_{0}[\hat{C},C;\hat{\bb},\bb]$ is the value of the
action at the stationary point: 
\begin{align}
\frac{\delta}{\delta C^{ab}}{\cal L}[\hat{C},C&;\hat{\bb},\bb]
  = 0 
   \nonumber\\
 & \quad\Rightarrow\quad 
     i\hat{C}^{ab}=\frac{1}{N}\sum_{i}\langle i\hat{h}_{i}^{a}i\hat{h}_{i}^{b}\rangle_{0},
\end{align}
\begin{align}
    \frac{\delta}{\delta i\hat{C}^{ab}}{\cal L}&[\hat{C},C;\hat{\bb},\bb] 
    = 0 
   \nonumber \\
    &\quad\Rightarrow\quad 
    C^{ab}=\frac{1}{N}\sum_{i}\langle S_{i}^{a}S_{i}^{b}\rangle_{0}+\epsilon i\hat{C}^{ab}.\label{eq:C_ab}
\end{align}
The (self-consistent) average $\langle(\dotso)\rangle_{0}$ is over
all paths of the dynamical process $\{\hat{h},h\}_{t\in[t_{0},t]}$
governed by the action ${\cal L}[\hat{h},h;\hat{C},C,\hat{\bb},\bb]$
evaluated at the stationary point. The normalisation $\overline{Z}[0,\bb]=1$
implies that $\hat{C}^{ab}=0$ is the correct self-consistent solution,
see below. Then $\overline{Z}_{0}[\hat{\bb},\bb]=\prod_{i}Z_{i}[\hat{\bb},\bb]$
with 
\begin{equation}
Z_{i}[\hat{\bb},\bb]=\int{\cal D}h_{i}\,{\cal D}\hat{h}_{i}\,e^{-S[\hat{h}_{i},h_{i};0,C]+\sum_{a}(i\hat{h}_{i}^{a}\bb_{i}^{a}+i\hat{\bb}_{i}^{a}h_{i}^{a})},
\end{equation}
and the dynamical behaviour of the network in the limit $N\to\infty$
is fully described by the single-site dynamical processes $\{\hat{h}_{i},h_{i}\}$.

Using the identity 
\begin{equation}
\exp\left[\frac{1}{2}\sum_{ab}i\hat{h}_{i}^{a}C^{ab}i\hat{h}_{i}^{b}\right]=\left\langle \exp{\sum_{a}i\hat{h}_{i}^{a}\eta^{a}}\right\rangle _{\eta}
\end{equation}
where $\eta^{a}$ is Gaussian field of mean $\langle\eta^{a}\rangle_{\eta}=0$
and 
\begin{equation}
\langle\eta^{a}\eta^{b}\rangle_{\eta}=C^{ab},
\label{eq:scVar}
\end{equation}
$Z_{i}[\hat{\bb},\bb]$ can be written as 
\begin{widetext}
\begin{equation}
Z_{i}[\hat{\bb},\bb]=\left\langle \int{\mathcal{D}}\hat{h}_{i}\,{\mathcal{D}}h_{i}\,e^{-\sum_{a}i\hat{h}_{i}^{a}\bigl[(1+\partial_{a})h_{i}^{a}-\eta^{a}-h_{i}^{0}\delta_{a0}\bigr]+\sum_{a}(i\hat{h}_{i}^{a}\bb_{i}^{a}+i\hat{\bb}_{i}^{a}h_{i}^{a})}\right\rangle _{\eta}.
\end{equation}
\end{widetext}
Therefore, $Z_{i}[\hat{\bb},\bb]$ is the generating functional of
the stochastic process described by the stochastic differential equation,
\begin{equation}
\partial_{a}h_{i}^{a}=-h_{i}^{a}+\bb_{i}^{a}+\eta^{a},\label{eq:mfd}
\end{equation}
with initial condition $h_{i}(t_{0})=h_{i}^{0}$ averaged over the
Gaussian noise $\eta^{a}$. This process provides the full description
of the dynamics of the network in the limit $N\to\infty$. While diagonal
in the site index, the process maintains memory of the other sites
through the Gaussian field $\eta^{a}$ because $\langle\eta^{a}\eta^{b}\rangle_{\eta}$
must be computed self-consistently through the constraint \eqref{eq:C_ab}.
Equations \eqref{eq:C_ab}, \eqref{eq:scVar} and  \eqref{eq:mfd} are the central equations of the 
DMFT.

Note that if $\hat{C}^{ab}$ were not zero then $Z_{i}[0,\bb]=\langle\exp{[\sum_{ab}i\hat{C}^{ab}S_{i}^{a}S_{i}^{b}}]\rangle$,
where the average is over the stochastic process \eqref{eq:mfd}, and $\overline{Z}[0,\bb]$
would not be necessarily equal to $1$.

\section{\label{sec:DMFT_Sol}Solution of DMFT Equations.}

In this Section we discuss the solutions of the  DMFT
equations. Without loss of generality we can take uniform $\hat{\bb}^{a}$
and $\bb^{a}$ and drop the site index. Moreover, since we are interested
in the steady state, we take the initial time $t_{0}=-\infty$ and
we can neglect the initial state $h_{i}(t_{0})$.

\subsection{DMFT equations}
To discuss the DMFT it is useful to rewrite the DMFT equations as follows.
Using the relation $S^a = \phi(gh^a)$  the DMFT equations  \eqref{eq:C_ab} and  \eqref{eq:scVar} can 
be reduced to:
\begin{equation}
       \langle\eta^{a}\eta^{b}\rangle_{\eta}=C^{ab}=\langle\phi(gh^{a})\phi(gh^{b})\rangle_{\eta},
\label{eq:C_ab_D}
\end{equation}
where, from Eq. \eqref{eq:mfd}:
\begin{equation}
h^{a}=h(t_{a})=\int_{-\infty}^{t_{a}}dt_{b}\,e^{-(t_{a}-t_{b})}\,\eta(t_{b}).
\end{equation}
We have set $\bb_i^{a}=0$ because no external field is present in the original
problem.

The synaptic field $h^{a}$ is a linear functional of $\eta^{a}$,
and hence it is a Gaussian process with zero mean and correlation
\begin{equation}
\langle h^{a}h^{b}\rangle_{\eta}=\Delta^{ab}.
\end{equation}

Equation \eqref{eq:C_ab_D} provides a nonlinear relation between
the field correlation $\Delta^{ab}$ and
the activity correlation $C^{ab}$.
Explicitly, 
\begin{equation}
 C^{ab}=\int\int\frac{d^{2}\boldsymbol{\text{h}}}{2\pi\sqrt{\det\Delta}}\exp\left[-\frac{1} {2}\boldsymbol{\text{h}}^{T}\Delta^{-1}\boldsymbol{\text{\text{h}}}\right]\phi(gh^{a})\phi(gh^{b})\label{eq:CvsDel}
\end{equation}
where $\boldsymbol{\text{h}}^{T}=(h^{a},h^{b})$, and $\boldsymbol{\Delta}$
is the 2x2 symmetric matrix: 
\begin{equation}
   \boldsymbol{\Delta}=\begin{bmatrix}\Delta^{aa} & \Delta^{ab}\\
\Delta^{ab} & \Delta^{bb}
\end{bmatrix}.
\end{equation}

It is sometime convenient to write this relation as:
\begin{equation}
\label{eq:C_DD0-1}
 \begin{split}
    C^{ab}=\int & Dz  
     \int Dx\,\phi\Bigl(gx\sqrt{\Delta^{aa}-|\Delta^{ab}|}+gz\sqrt{|\Delta^{ab}|}\Bigr)
     \\ 
     & \times \int Dy\,\phi\Bigl(gy\sqrt{\Delta^{bb}-|\Delta^{ab}|}+gz\sqrt{|\Delta^{ab}|}\Bigr),
\end{split}     
\end{equation}
where $Dx=dx\exp(-x^{2}/2)/\sqrt{2\pi}$, and similarly $Dy$ and $Dz$,  are Gaussian measures.
Alternatively, introducing the Fourier transform $\tilde{\phi}(k)$  of the function
$\phi(x)$, the relation between $\Delta^{ab}$ and $C^{ab}$  can also be 
written as,
\be
 \begin{split}
   C^{ab}  = \int&\frac{dk}{2\pi}\frac{dk'}{2\pi}\ \tilde{\phi}(k)\,\tilde{\phi}(k')
    \\ 
    &\times 
    \exp\Bigl[-\frac{g^{2}}{2}
   (\Delta^{aa}k^{2}+\Delta^{bb}k'^{2})-g^{2}\Delta^{ab}kk'\Bigr],
\end{split}   
\label{eq:fourier_sol_r}
\ee
Details are in Appendix \ref{app:eq_class_pot}.

On the other hand, by multiplying  Eq. \eqref{eq:mfd} by itself and averaging over $\eta$,
we obtain the relation,
\begin{equation}
\label{eq:eq_D_ab}
(1+\partial_{a})(1+\partial_{b})\Delta^{ab}=C^{ab},
\end{equation}
expressing $\Delta^{ab}$ as function of  $C^{ab}$.

Equations \eqref{eq:CvsDel} and \eqref{eq:eq_D_ab} constitute the DMFT self-consistent 
equations of our system.

\subsection{Steady State Solutions}

The DMFT considerably simplifies in the steady state regime, which
is the focus of the present paper. In this regime, the dynamical correlation
functions are time translation invariant and $\Delta^{ab}$ depends
on $t_{a}$ and $t_{b}$ only through the time difference $\tau=t_{a}-t_{b}$:
\begin{align}
\Delta^{ab}& =\Delta(\tau) \equiv \Delta,
\\
\Delta^{aa}&=\Delta^{bb} = \Delta(0) \equiv\Delta_{0},
\end{align}
In this case the DMFT equation \eqref{eq:C_DD0-1} becomes: 
\begin{equation}
\label{eq:C_DD0}
   C(\Delta;\Delta_{0})=\int Dz\left[\int Dx\,\phi\Bigl(gx\sqrt{\Delta_{0}-|\Delta|}+gz\sqrt{|\Delta|}\Bigr)
   \right]^{2},
\end{equation}
while,
using the identities 
$\partial_a \Delta(t_a-t_b) = \partial_\tau \Delta(\tau)$
and
 $\partial_b \Delta(t_a-t_b) = -\partial_\tau \Delta(\tau)$,
Eq. \eqref{eq:eq_D_ab} reduces to 
\begin{equation}
\label{eq:eq_D}
\Delta-\partial_{\tau}^{2}\Delta=C(\Delta;\Delta_{0}).
\end{equation}
Since $\Delta(\tau)$ is a correlation function, acceptable solutions to Eq. \eqref{eq:eq_D} must obey 
\be
\abs{\Delta(\tau)}\leq\Delta(0),
\ee
and in particular they must be bounded. 

Equation \eqref{eq:eq_D} admits a two-parameter family of solutions 
parametrised by the \textit{initial conditions} $\Delta(0)$ and  $\partial_{\tau}\Delta|_{\tau=0}=0$.
This choice is rather convenient because
the initial ``velocity'' is 
\begin{equation}
\partial_{\tau}\Delta|_{\tau=0}=0,
\end{equation}
as follows from the explicit solution of the differential equation \eqref{eq:eq_D}
\begin{equation}
\Delta(\tau)=\frac{1}{2}\int_{-\infty}^{+\infty}d\tau'\,e^{-|\tau-\tau'|}\,C(\tau'),
\end{equation}
which implies that $\Delta(\tau)$ is a differentiable even function
of $\tau$: $\Delta(-\tau)=\Delta(\tau)$. 
The initial ``position'' $\Delta(0)$ is fixed by the requirement
\begin{equation}
\Delta(0) = \Delta_{0},
\end{equation}
so that the steady state DMFT solutions are a one-parameter family of curve
$\Delta \equiv \Delta(\tau; \Delta_0)$ parameterised by  $\Delta_0$.

\subsubsection{Fixed Point Solution}

The simplest solution is that of a fixed point:  $\Delta(\tau)=\Delta_0=C$,
leading to the  self consistent equation,

\begin{equation}
\Delta_0=[\phi^{2}]_{\Delta_0},
\end{equation}
where, for later use, we have introduced the notation
\begin{equation}
[f]_{\Delta_0}=\int Dx\ f\Bigl(gx\sqrt{\Delta_0}\Bigr).
\end{equation}
%
For the odd-symmetric transfer function, such as $\phi(x)=\tanh(x)$, there
is always a solution with $\Delta_0=0$, corresponding to the zero fixed
point $h_{i}=0$ of the original dynamics. 
A solution with nonzero $\Delta_0$ appears when $g>1$ . 
The static fixed point solution is however unstable for $g>1$, as shown in the next Section,
see also Appendix \ref{app:Amari}. 

\subsubsection{Time-dependent solution: Potential and Energy}

Solving the DMFT equations is greatly facilitated by noting that for a fixed
$\Delta_{0}$ the differential equation \eqref{eq:eq_D} can be viewed as the inertial
dynamics of a particle moving under the influence of a potential $V(\Delta;\Delta_{0}),$
i.e.,

\begin{equation}
\partial_{\tau}^{2}\Delta=-\partial_{\Delta}V(\Delta;\Delta_{0}),\label{eq:Newton_eq}
\end{equation}
where, 
\begin{equation}
V(\Delta;\Delta_{0})=-\frac{\Delta^{2}}{2}+\int_{0}^{\Delta}d\Delta'\,C(\Delta';\Delta_{0}).
\end{equation}
Introducing the function $\Phi(x)=\int_{0}^{x}dy\,\phi(y)$, primitive
of the gain function $\phi(x)$, the potential can be expressed as:
\be
\begin{split}V(\Delta;&\Delta_{0})  =  -\frac{\Delta^{2}}{2}
\\
 & +\frac{1}{g^{2}}\int Dz\left[\int Dx\,\Phi\Bigl(gx\sqrt{\Delta_{0}-|\Delta|}+gz\sqrt{|\Delta|}\Bigr)\right]^{2}\\
 & \phantom{====}-\frac{1}{g^{2}}\left[\int Dx\,\Phi\Bigl(gx\sqrt{\Delta_{0}}\Bigr)\right]^{2},
\end{split}
\label{eq:class_pot}
\ee
The last term ensures that $V(0,\Delta_0)=0$.
Details can be found in Appendix \ref{app:eq_class_pot}. All solutions to 
Eq. \eqref{eq:Newton_eq} conserve the energy:
\begin{equation}
E_{{\rm c}}=\frac{1}{2}\bigl(\partial_{\tau}\Delta\bigr)^{2}+V(\Delta;\Delta_{0}).
\label{eq:Ec}
\end{equation}
Hence, since DMFT solutions must have $\Delta(0)=\Delta_0$, $\partial_{\tau}\Delta|_{\tau=0}=0$
and be bounded,
all solutions to Eq. \eqref{eq:Newton_eq} with energy $E_{{\rm c}}=V(\Delta_{0};\Delta_{0})$
leading to bounded  orbits are possible DMFT solutions $\Delta=\Delta(\tau;\Delta_0)$.
The qualitative behaviour of the solutions can be inferred from the shape of the
potential $V(\Delta; \Delta_0)$.
Solutions with different properties are possible because $V(\Delta; \Delta_0)$ depends 
parametrically on the value of $\Delta_0$, reflecting the self-consistent nature of the 
DMFT.

\subsubsection{Phase Diagram}

The exact form of $V(\Delta;\Delta_{0})$ depends on $\phi(x)$. However,
its qualitative behaviour can be determined.
First, we note that for $\Delta>0$:
\begin{widetext}
\begin{equation}
\partial_{\Delta}^{3}V(\Delta;\Delta_{0})=g^{4}\int Dz\left[\int Dx\,\phi''\Bigl(gx\sqrt{\Delta_{0}-\Delta}+gz\sqrt{\Delta}\Bigr)\right]^{2}>0.
\end{equation}
\end{widetext}
The ``prime'' denotes derivative of the function with respect to
its argument, hence $\phi''(x)$ is the second derivative of $\phi(x)$
with respect to $x$. Thus $\partial_{\Delta}^{2}V(\Delta;\Delta_{0})$
is monotonously increasing and can vanish at most once for $0<\Delta<\Delta_{0}$.
As a consequence, the shape of $V(\Delta;\Delta_{0})$ is either a
single-well or a double-well depending on the sign of $\partial_{\Delta}^{2}V(\Delta;\Delta_{0})$
at $\Delta=0$. Expanding $V(\Delta;\Delta_{0})$ for $\abs{\Delta}\ll1$,
gives, see \ref{app:eq_class_pot}, 
\begin{equation}
V(\Delta;\Delta_{0})=\Bigl(-1+g^{2}[\phi']_{\Delta_{0}}^{2}\Bigr)\frac{\Delta^{2}}{2}+g^{6}[\phi''']_{\Delta_{o}}^{2}\frac{\Delta^{4}}{24}+\dotsc.
\end{equation}
If $-1+g^{2}[\phi']_{\Delta_{0}}\geq0$ then the potential is a single
well: the energy $E_{{\rm c}}$ is positive and $\Delta(\tau)$ is
time-periodic. It changes sign during one oscillation.

\begin{figure}[tbh]
\centering \includegraphics[scale=0.65]{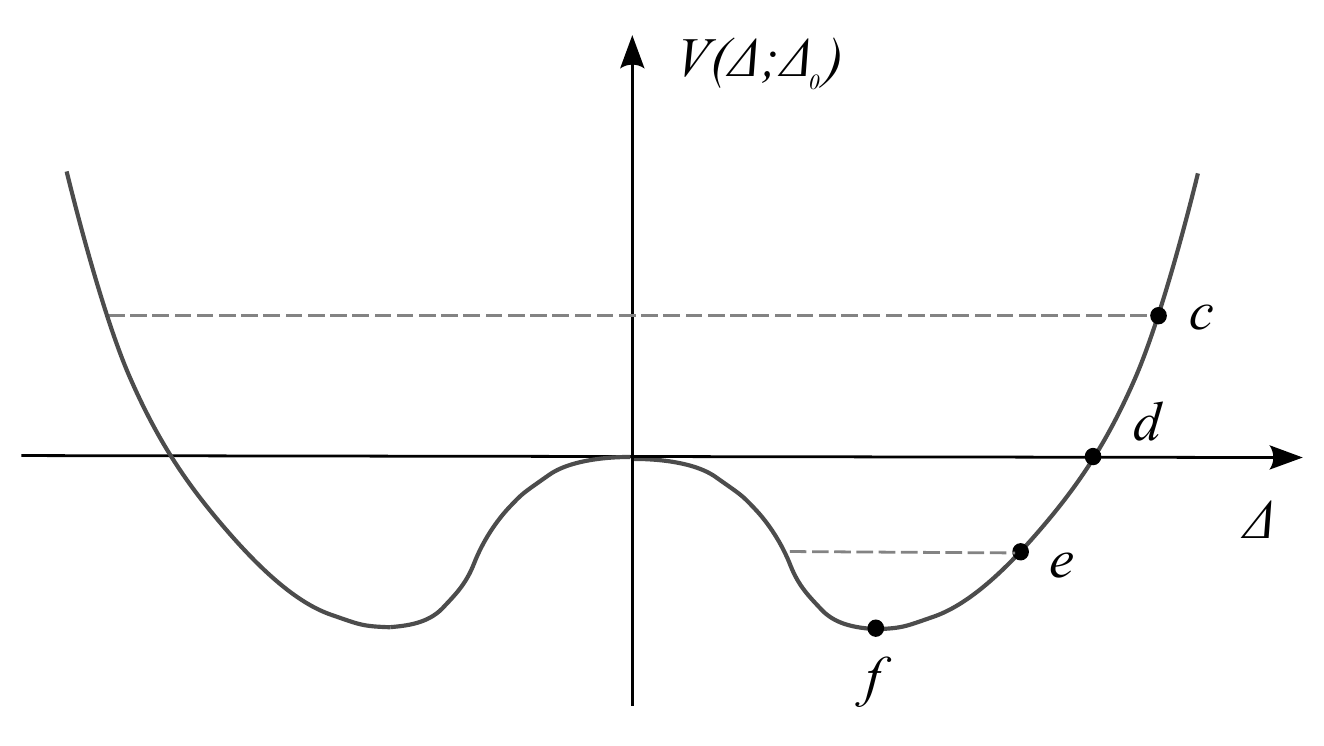} 
\caption{ Qualitative behaviour of $V(\Delta;\Delta_{0})$ for $-1+g^{2}[\phi']_{\Delta_{0}}<0$.
Labels denotes the different possible behaviours of the solution.
Label $c$: $E_{{\rm c}}>0$, $\Delta(\tau)$ is periodic and changes
sign. Label $e$: $E_{{\rm c}}<0$, $\Delta(\tau)$ is periodic but
remains positive. Label $d$: $E_{{\rm c}}=0$, $\Delta(\tau)$ decays
to zero as $\tau\to\infty$. Label $f$: minimum allowable value of
$E_{{\rm c}}$, $\Delta(\tau)=\Delta_{0}$, static solution. }
\label{fig:CPot} 
\end{figure}

In the case $-1+g^{2}[\phi']_{\Delta_{0}}<0$ the potential has a
double-well shape, and qualitatively different solutions appears depending
on the sign of the energy 
\begin{equation}
E_{{\rm c}}=V(\Delta_{0};\Delta_{0})=-\frac{\Delta_{0}^{2}}{2}+\frac{1}{g^{2}}[\Phi^{2}]_{\Delta_{0}}-\frac{1}{g^{2}}[\Phi]_{\Delta_{0}}^{2}.
\end{equation}
When $E_{{\rm c}}>0$ the solution is qualitatively similar to the
previous case: $\Delta(\tau)$ is time-periodic and changes sign during
one oscillation. On the contrary if $E_{{\rm c}}<0$ and $\partial_{\Delta}V(\Delta;\Delta_{0})$
at $\Delta=\Delta_{0}$ is positive then $\Delta(\tau)$ is time-periodic
but does not change sign during one oscillation. The two regimes are
separated by the boundary $E_{{\rm c}}=0$, where $\Delta(\tau)$
decays monotonously to $0$ as $\tau\to\infty$. When $E_{{\rm c}}$
reaches the minimum of $V(\Delta;\Delta_{0})$, the solution becomes
\textit{time-independent}. This occurs for, 
\begin{equation}
\partial_{\Delta}V(\Delta;\Delta_{0})\Bigl|_{\Delta=\Delta_{0}}=-\Delta_{0}+C(\Delta_{0};\Delta_{0})=0,
\end{equation}
and one recovers the time-independent solution found previously. The
different cases are shown in Fig. \ref{fig:CPot}. 

\begin{figure}[tbh]
\centering \includegraphics[scale=0.5]{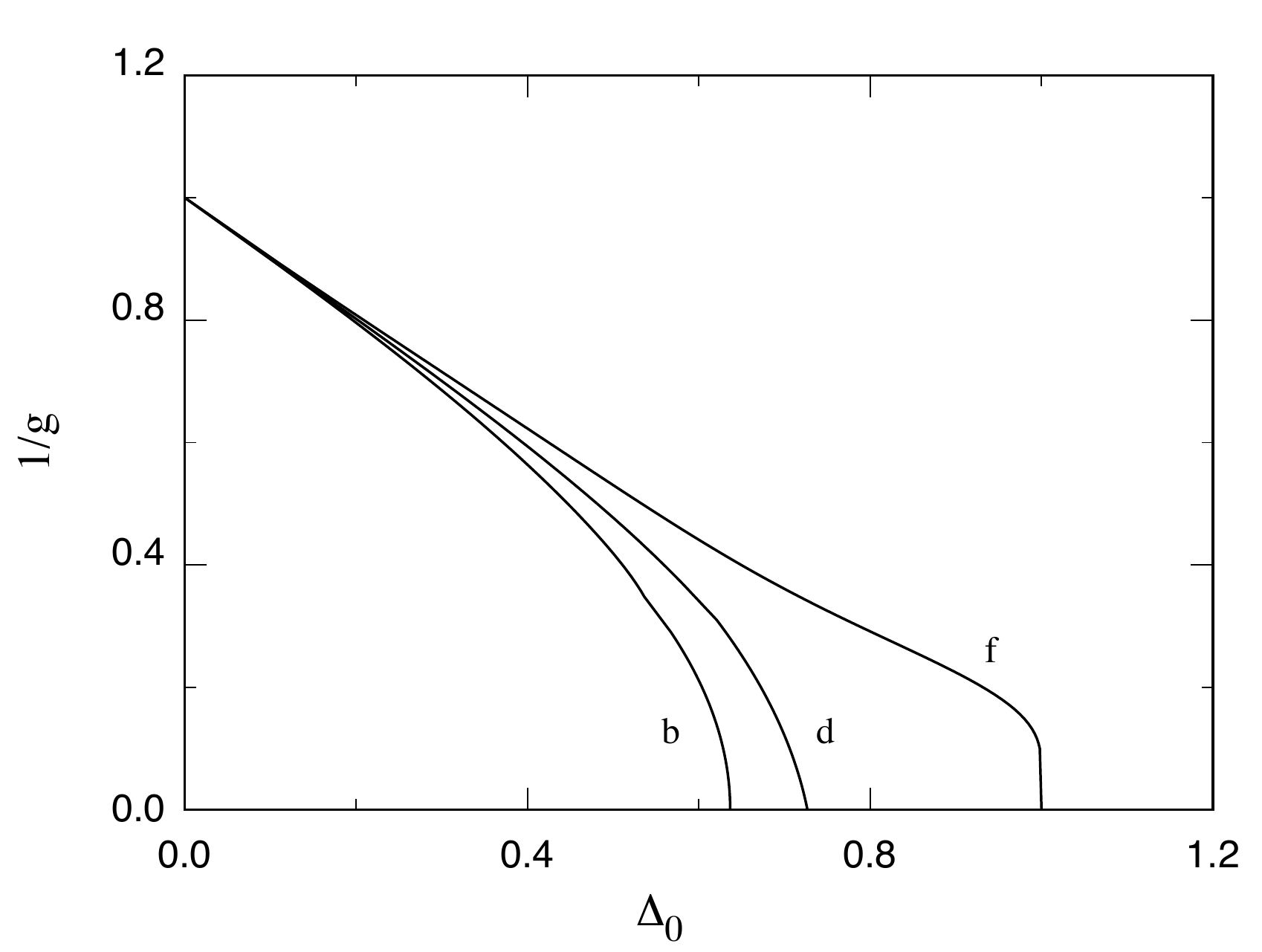} \caption{ Dynamical mean field theory phase diagram. The curves delimit the
regions of qualitative different behaviours. On the curve $f$ the
energy $E_{{\rm c}}$ is equal to the minimum of $V(\Delta;\Delta_{0})$
and $\Delta(\tau)$ is time-independent. Between the curve $f$ and
the curve $d$ the energy $E_{{\rm c}}<0$ and $\Delta(\tau)$ is
time-periodic but positive. On the curve $d$ the energy $E_{{\rm c}}$
vanishes and $\Delta(\tau)$ decays to $0$ as $\tau\to\infty$. Below
the curve $d$ the energy $E_{{\rm c}}>0$ and $\Delta(\tau)$ is
time-periodic with not definite sign. On the curve $b$ the potential
$V(\Delta;\Delta_{0})$ changes from a double well shape to a single
well shape. Above the curve $f$ there are no solution to the DMFT
equations. For $g>1$ all curves collapse and only the
static solution $\Delta(\tau)=\Delta_{0}=0$ survives. Numerical values
are for $\phi(x)=\tanh(x)$. }
\label{fig:PDiag} 
\end{figure}

By drawing in the plane $(\Delta_{0},1/g)$ the curves separating the different type of solutions  we obtain the phase diagram shown in
Fig. \ref{fig:PDiag}. 


Above the curve $f$ there are no solutions with $\Delta_{0}>0$.
Thus for $g<1$ only the time-independent solution $\Delta=\Delta_{0}=0$
exists. The vanishing of the equal time correlation $\Delta_{0}$
in the steady state implies that the system flows to the zero fix
point $h_{i}=0$. The stability of this solution for $g<1$ can be
deduced by linearising Eq. \eqref{eq:kirch} and noting that the maximum
real part of the eigenvalues of the matrix $J_{ij}$ is $1$.

For $g>1$ different scenarios appears. On the curve $f$ the energy
$E_{{\rm c}}$ attains its minimum value and the solution is time-independent:
$\Delta(\tau)=\Delta_{0}$. On this curve the state the network flows
to a non-zero fix-point characterised by a non-trivial distribution
of $h_{i}$. In the region below the curve $f$ the energy $E_{{\rm c}}$
is larger than the minimum of $V(\Delta;\Delta_{0})$ and $\Delta(\tau)$
becomes time-dependent. Here time-periodic solutions appear, either
changing sign in one period, below curve $d$, or not, between curves
$d$ and $f$. In either cases these solutions imply that the dynamical
behaviour of the network in the steady state is a limit cycle. On
the curve $d$, corresponding to $E_{{\rm c}}=0$ and separating the
two types of periodic solutions, $\Delta(\tau)$ is not periodic and
decays monotonously to $0$ as $\tau\to\infty$. On this curve the
dynamical behaviour of the network is chaotic.

\section{\label{sec:DMFT_Stab}Fluctuations around the DMFT and solution Selection}

The large number of solutions of the DMFT for
$g>1$ raises the question of what are the criteria of selection of
one or a few of those solutions as the physically relevant ones. The
problem is twofold. The DMFT follows from a
saddle point calculation of the path integrals. Thus only solutions
leading to a stable saddle point, i.e., a local maximum of the action,
must be retained. An analysis of the Hessian of the fluctuations reveals
that all solutions are stable saddle point (see below); the
steady state of the network is hence given by all the above mentioned
solutions: fixed points, limit cycles and chaos.

The second problem is which of these solutions represent a stable
attractor of the network dynamics. Stated differently, towards which
steady state the dynamics will flow with probability one as the system
size $N\to\infty$? We address this question by studying the behaviour
of two copies (replicas) of the network as $N\to\infty$.

\subsection{\label{sec:RDMFT}Two replica formalism}

The stability of an attractor is related to the properties of the
linear response matrix 
$\langle\partial h_{i}(t+\tau)/\partial h_{j}(t)\rangle_{J}\sim\langle h_{i}(t+\tau)\hat{h}{}_{j}(t)\rangle_{J}$.
Due to the quenched random couplings $J_{ij}$ instability in this matrix may be
washed out by averaging over $J_{ij}$, hence to uncover instability, one
needs to consider quantities such as $\overline{\langle h_{i}(t+\tau)\hat{h}{}_{j}(t)\rangle_{J}^{2}}$.
Such averages can be computed using DMFT starting from two identical
copies of the system, namely, $h_{i}^{\alpha}(t),\:\alpha=1,2$ ,
obeying 
\be
\label{eq:replica}
\frac{d}{dt}\,h_{i}^{\alpha}(t)=-h_{i}^{\alpha}(t)+\sum_{j=1}^{N}J_{ij}\,\phi(gh^{\alpha}(t)),
\quad i=1,\dotsc,N,
\ee
and evaluating 
\be
\overline{\langle h_{i}(t+\tau)\hat{h}{}_{j}(t)\rangle_{J}^{2}}
=\overline{\langle h_{i}^{1}(t+\tau)h_{i}^{2}(t'+\tau)\hat{h}{}_{j}^{1}(t)\hat{h}{}_{j}^{2}(t')\rangle_{J}}
\ee
with $t\not=t'$. Hence the full dynamic stability can be determined
from a stability analysis of the path integral formulation of the
replicated system \eqref{eq:replica}.

Conveniently, our above results incorporates readily the replicated
system, if we replace the index $a$ in Eq. \eqref{eq:delta_id} by $\bs{a}=(\alpha,a)=(\alpha,t_{a})$
representing both replica index and time, e.g., 
$h_{i}^{\bs{a}}=h_{i}^{\alpha,a}=h_{i}^{\alpha}(t_{a})$.
For example, the averaged generating functional $\overline{Z}[\hat{\bb},\bb]$
of the replicated system can be read directly from 
Eqs. \eqref{eq:gen_fun_ave}-\eqref{eq:DFT_hh_S}.
In particular the action can be written as 
\begin{equation}
 \label{eq:DFT_S_r}
   {\cal L}[\hat{C},C;\hat{\bb},\bb]
       =\frac{1}{2}\sum_{\bs{ab}}\bigl(i\hat{C}^{\bs{ab}}C^{\bs{ab}}
        + \frac{\epsilon}{2}\,\hat{C}^{\bs{ab}}\hat{C}^{\bs{ab}}\bigr)
        -W[\hat{C},C;\hat{\bb},\bb],
\end{equation}
where 
\begin{equation}
\begin{split}
NW[\hat{C},C;\hat{\bb},\bb]=\sum_{i}&\ln\int{\cal D}\hat{h}_{i}\,{\cal D}h_{i}
\\
&\times 
e^{-S[\hat{h}_{i},h_{i};C,\hat{C}]+\sum_{\bs{a}}(i\hat{h}_{i}^{\bs{a}}\bb_{i}^{\bs{a}}+i\hat{\bb}_{i}^{\bs{a}}h_{i}^{\bs{a}})},
\end{split}
\end{equation}
and 
\begin{equation}
\begin{split}S[\hat{h}_{i},h_{i};\hat{C},C] =
      &\sum_{\bs{a}}\Bigl[i\hat{h}_{i}^{\bs{a}}(1+  \partial_{a})h_{i}^{\bs{a}}
                -h_{i}^{0,\bs{a}}\delta_{a0}\Bigr]\\
 & -\frac{1}{2}\sum_{\bs{a}\bs{b}}\Bigl[i\hat{C}^{\bs{a}\bs{b}}S_{i}^{\bs{a}}S_{i}^{\bs{b}}
  +C^{\bs{a}\bs{b}}i\hat{h}_{i}^{\bs{a}}i\hat{h}_{i}^{\bs{b}}\Bigr],
\end{split}
\label{eq:DFT_hh_S_r}
\end{equation}
where $\sum_{\bs{a}}\equiv\sum_{\alpha=1,2}\int dt_{a}$,
$C^{\bs{ab}}=C^{\alpha\beta,ab}=C^{\alpha\beta}(t_{a},t_{b})$
and similarly, 
$\hat{C}^{\bs{ab}}=\hat{C}^{\alpha\beta,ab}=\hat{C}^{\alpha\beta}(t_{a},t_{b})$.
As a consequence, the results of Sec. \ref{sec:DFT_G} can be immediately
extended to the replicated system. Hence, in the limit $N\to\infty$
the behaviour of the replicated
system is described by the saddle point of the functional \eqref{eq:DFT_S_r},
leading to the $2$-replica DMFT. 
At the saddle point $\hat{C}^{\bs{ab}}=0$
while $C^{\bs{ab}}$, solution of the stationary point equations,
may in general depend on both replica and time indexes. However, since
the two copies of the system are identical (including their initial
value), the DMFT order parameters cannot depend on the replica index
and hence $C^{\bs{ab}}=C^{ab}=C(t_{a},t_{b})$ for all $\alpha,\beta$.

\subsubsection{Fluctuations of the replicated DMFT:}

The stability of the $2$-replica DMFT solutions can be inferred from
the analysis of the Gaussian fluctuations about the stationary point
of the action ${\cal L}[\hat{C},C;\hat{\bb};\bb]$ of the replicated
system. Denoting by $Q^{\bs{ab}}$ and $\hat{Q}^{\bs{ab}}$
the fluctuations and expanding the action \eqref{eq:DFT_S_r} to second
order in $Q$ and $\hat{Q}$ leads to: 
\begin{equation}
      \overline{Z}[\hat{\bb},\bb] \sim \overline{Z}_{0}[\hat{\bb},\bb]
             \int{\cal D}\hat{Q}\,{\cal D}Q\,e^{-N{\cal L}_{2}[\hat{Q},Q;\hat{\bb},\bb]}, 
             \qquad N\gg1,
\label{eq:Z_fluct}
\end{equation}
where 
\begin{equation}
\begin{split}{\cal L}_{2}[\hat{Q},Q;\hat{\bb},\bb]=
  \frac{1}{2}&\sum_{\bs{ab}}  i\hat{Q}^{\bs{ab}}Q^{\bs{ab}}
  \\
  &-\frac{1}{8}  \sum_{\bs{ab},\bs{cd}}i\hat{Q}^{\bs{ab}}{\cal M}^{\bs{ab};\bs{cd}}i\hat{Q}^{\bs{cd}}
  \\
 & -\frac{1}{4}\sum_{\bs{ab},\bs{cd}}i\hat{Q}^{\bs{ab}}\bigl\langle S^{\bs{a}}S^{\bs{b}}i\hat{h}^{\bs{c}}
 i\hat{h}^{\bs{d}}\bigr\rangle_{0}Q^{\bs{cd}},
\end{split}
\end{equation}
with 
\begin{equation}
{\cal M}^{\mathbf{ab};\mathbf{cd}}=  \epsilon\, \delta_{\bs{ab},\bs{cd}} +
              \langle S^{\mathbf{a}}S^{\mathbf{b}}S^{\mathbf{c}}S^{\mathbf{d}}\rangle_{0}
      -\langle S^{\mathbf{a}}S^{\mathbf{b}}\rangle_{0}\langle S^{\mathbf{c}}S^{\mathbf{d}}\rangle_{0},
\end{equation}
and
\begin{equation}
\delta_{\bs{ab,cd}}=\delta_{\bs{ac}}\delta_{\bs{bd}}+\delta_{\bs{ad}}\delta_{\bs{bc}},
\end{equation}
is the symmetrized $\delta$-function.
The average $\langle(\dotso)\rangle_{0}$ is over the dynamical
process governed by the action 
\eqref{eq:DFT_hh_S_r}  with $\hat{C}^{\boldsymbol{\text{ab}}}$
and $C^{\boldsymbol{\text{ab}}}$ evaluated at the stationary point
of ${\cal L}[\hat{C},C;\hat{\bb};\bb]$.

Using the identity 
\begin{equation}
e^{-S[\hat{h},h;\hat{C},C]}\,i\hat{h}^{\bs{a}}i\hat{h}^{\bs{b}}
      =\frac{\delta}{\delta C^{\bs{ab}}}\,e^{-S[\hat{h},h;\hat{C},C]},
\end{equation}
the average $\langle S^{\bs{a}}S^{\bs{b}}i\hat{h}^{\bs{c}}i\hat{h}^{\bs{d}}\rangle_{0}$
is equal to: 
\begin{equation}
\bigl\langle S^{\bs{a}}S^{\bs{b}}i\hat{h}^{\bs{c}}i\hat{h}^{\bs{d}}\bigr\rangle_{0}=
      \frac{\delta}{\delta C^{\bs{cd}}}\,\bigl\langle S^{\bs{a}}S^{\bs{b}}\bigr\rangle_{0}.
\end{equation}
The derivative is evaluated by recalling that 
$\langle S^{\bs{a}}S^{\bs{b}}\rangle_{0}=\,\bigl\langle\phi(gh^{\bs{a}})\phi(gh^{\bs{b}})\bigr\rangle_{0}$,
where $h^{\bs{a}}$ is the solution of the DMFT stochastic differential equation
\begin{equation}
\partial_{a}h^{\bs{a}}=-h^{\bs{a}}+\eta^{\bs{a}},
\end{equation}
with $\eta^{\bs{a}}$ Gaussian field of zero mean and variance
$\langle \eta^{\bs{a}}\eta^{\bs{b}}\rangle_\eta = C^{\bs{ab}}$, cf. Sec. \ref{sec:DMFT}.
The average $\langle S^{\bs{a}}S^{\bs{b}}\rangle_{0}$ is thus a function of the field-field 
correlation function $\Delta^{\bs{ab}}=\langle h^{\bs{a}}h^{\bs{b}}\rangle_{\eta}$,
so that using the chain rule
$\langle S^{\bs{a}}S^{\bs{b}}i\hat{h}^{\bs{c}}i\hat{h}^{\bs{d}}\rangle_{0}$ is given by:
\begin{equation}
\bigl\langle S^{\bs{a}}S^{\bs{b}}i\hat{h}^{\bs{c}}i\hat{h}^{\bs{d}}\bigr\rangle_{0}=
             \frac{\partial}{\partial\Delta^{\bs{cd}}}\bigl\langle S^{\bs{a}}S^{\bs{b}}\bigr\rangle_{0}\,
             \frac{\delta\Delta^{\bs{cd}}}{\delta C^{\bs{cd}}},
\end{equation}
where, from the DMFT equations, $\delta\Delta^{\bs{ab}}/\delta C^{\bs{cd}}$ is solution of
\begin{equation}
     (1+\partial_{a})(1+\partial_{b})\, \frac{\delta\Delta^{\bs{ab}}}{\delta C^{\bs{cd}}}=
            \delta_{\bs{ac},\bs{bd}}.
\end{equation}

To further proceed, it is then more convenient to transform ${\cal L}_{2}$ to the equivalent
quadratic form: 
\begin{equation}
\begin{split}
     {\cal L}_{2}[\hat{Q},\Psi;\hat{\bb},\bb] = &
            -\frac{1}{8}\sum_{\bs{ab},\bs{cd}}i\hat{Q}^{\bs{ab}}{\cal M}^{\bs{ab};\bs{cd}}i\hat{Q}^{\bs{cd}}
            \\
           & +\frac{1}{4} \sum_{\bs{ab},\bs{cd}}i\hat{Q}^{\bs{ab}}{\cal A}^{\bs{ab};\bs{cd}}\Psi^{\bs{cd}},
\end{split}           
\label{eq:S2}
\end{equation}
where $\Psi^{\bs{ab}}$ is defined through, 
\begin{equation}
(1+\partial_{a})(1+\partial_{b})\Psi^{\bs{ab}}=Q^{\bs{ab}},
\end{equation}
and the operator ${\cal A}$ acting on $\Psi$ via, 
\begin{equation}
{\cal A}^{\bs{ab};\bs{cd}}:=(1+\partial_{a})(1+\partial_{b})\delta_{\bs{ac},\bs{bd}}
                    -\frac{\partial}{\partial\Delta^{\bs{cd}}}\bigl\langle S^{\bs{a}}S^{\bs{b}}\bigr\rangle_{0}.
 \label{eq:ep_H}
\end{equation}

The (Gaussian) integration over $\hat{Q}$ in Eq. \eqref{eq:Z_fluct}
is well defined and can be performed.
It leads to a term of the form $\exp[-(const)\Psi{\cal A}^{\dag}{\cal M}{}^{-1}{\cal A}\Psi]$,
where ${\cal A}^{\dag}$ is the adjoint of ${\cal A}$. Stability
of the stationary point requires that the operator ${\cal A}$ has
no zero eigenvalue. 

Making use of the explicit form of 
$\partial\bigl\langle S^{\bs{a}}S^{\bs{b}}\bigr\rangle_{0} / \partial\Delta^{\bs{cd}}$:
\begin{equation}
\begin{split}
     \frac{\partial}{\partial\Delta^{\bs{cd}}}\bigl\langle S^{\bs{a}}S^{\bs{b}}\bigr\rangle_{0}
         = & \frac{\partial}{\partial\Delta^{\bs{ab}}}\bigl\langle S^{\bs{a}}S^{\bs{b}}\bigr\rangle_{0}
         \delta_{\bs{ac}},\delta_{\bs{bd}}
         \\
        & +\frac{\partial}{\partial\Delta^{\bs{aa}}}\bigl\langle S^{\bs{a}}S^{\bs{b}}\bigr\rangle_{0}
            \delta_{\bs{ca}}\delta_{\bs{da}}
            \\
      &  +\frac{\partial}{\partial\Delta^{\bs{bb}}}\bigl\langle S^{\bs{a}}S^{\bs{b}}\bigr\rangle_{0}
        \delta_{\bs{cb}}\delta_{\bs{db}},
\end{split}        
\end{equation}
the eigenvalue equation for the operator ${\cal A}$ reads:
\begin{widetext}
\begin{equation}
   (1+\partial_{a})(1+\partial_{b})\Psi^{\bs{ab}}
    - \frac{\partial}{\partial\Delta^{\bs{ab}}}\bigl\langle S^{\bs{a}}S^{\bs{b}}\bigr\rangle_{0}\Psi^{\bs{ab}}
    - \frac{\partial}{\partial\Delta^{\bs{aa}}}\bigl\langle S^{\bs{a}}S^{\bs{b}}\bigr\rangle_{0} \Psi^{\bs{aa}}
    - \frac{\partial}{\partial\Delta^{\bs{bb}}}\bigl\langle S^{\bs{a}}S^{\bs{b}}\bigr\rangle_{0}\Psi^{\bs{bb}}
    = \Lambda \Psi^{\bs{ab}}.
\label{eq:Eigen_Eq}
\end{equation}
\end{widetext}
The stability condition requires that this equation must \textit{not}
admit a solution with $\Lambda=0$. 
The stability criterion does not require an evaluation of ${\cal M}$. 

Note that the intra-replica fluctuations $\alpha=\beta$ 
are decoupled and independent of the inter-replica fluctuations $\alpha\not=\beta$.
 Note also that since ${\cal A}$ is a symmetric operator,
the solutions to the eigenvalue equation \eqref{eq:Eigen_Eq} can
be classified as either symmetric eigenvectors  $\Psi^{\mathbf{ab}}=\Psi^{\mathbf{ba}}$
or antisymmetric eigenvectors $\Psi^{\mathbf{ab}}=-\Psi^{\mathbf{ab}}$, where
the symmetry operation is the simultaneous exchange of both replica
and time indices.

\subsection{Stability of the time-independent solution}

The general expression of time-independent DMFT solution 
$\Delta^{\alpha\beta}=C^{\alpha\beta}=\bigl\langle\phi(gh^{\alpha})\phi(gh^{\beta})\bigr\rangle_{0}$
can be written as in Eq. \eqref{eq:fourier_sol_r}:
\begin{equation}
\begin{split}
       \Delta^{\alpha\beta} =
    \int&\frac{dk}{2\pi}\frac{dk'}{2\pi}\ \tilde{\phi}(k)\,\tilde{\phi}(k')
    \\
    &\times
    \exp\Bigl[-\frac{g^{2}}{2}
   (\Delta^{\alpha\alpha}k^{2}+\Delta^{\beta\beta}k'^{2})-g^{2}\Delta^{\alpha\beta}kk'\Bigr],
\end{split}   
\end{equation}
where $\tilde{\phi}(k)$ is the Fourier transform of the function
$\phi(x)$. 

The relevant solution to these equations is $\Delta^{\alpha\beta} = \Delta$,
where $\Delta$ is obtained from the self-consistent equation:
\begin{equation}
\label{eq:t-ind-sol}
\Delta=\int\frac{d\eta}{\sqrt{2\pi}}\,e^{-\eta^{2}/2}\,\phi(g\sqrt{\Delta}\, h)^{2} =
\bigl[\phi^2\bigr]_{\Delta},
\end{equation}
as in the single replica time independent solution.
For this solution, using Eqs. \eqref{eq:AppC_der_aa}-\eqref{eq:AppC_der_bb},
$\partial \bigl\langle S^{\bs{a}}S^{\bs{b}}\bigr\rangle_{0} / \partial\Delta^{\bs{ab}}= 
g^2 [(\phi')^{2}]_{\Delta}$
and 
$ \partial \bigl\langle S^{\bs{a}}S^{\bs{b}}\bigr\rangle_{0} / \partial\Delta^{\bs{aa}}=
(g^{2}/2)[\phi\phi'']_{\Delta}$, and 
the eigenvalue equation \eqref{eq:Eigen_Eq} becomes: 
\begin{equation}
\begin{split}
        \Bigl[(1+\partial_{a})(1+&\partial_{b})-g^{2}[(\phi')^{2}]_{\Delta}\Bigr]\Psi^{\bs{ab}}
        \\
        &-\frac{g^{2}}{2}[\phi\phi'']_{\Delta}\Bigl[\Psi^{\bs{aa}}+\Psi^{\bs{bb}}\Bigr]=\Lambda\Psi^{\bs{ab}}.
\end{split}        
\label{eq:eig_eq_st}
\end{equation}

Since $\phi(0)=0$ equation \eqref{eq:t-ind-sol}  admits the trivial solution
$\Delta=0$ for all $g$.
In this case,  recalling that by assumption $\phi'(0)=1$,
the eigenvalue equation \eqref{eq:eig_eq_st} reduces to: 
\begin{equation}
\Bigl[(1+\partial_{a})(1+\partial_{b})-g^{2}\Bigr]\Psi^{\mathbf{ab}}=\Lambda\Psi^{\mathbf{ab}}.
\end{equation}
Taking the Fourier transform with respect to $t_{a}$ and $t_{b}$,
we find 
\begin{equation}
\Lambda=(1-i\omega_{a}+\delta)(1-i\omega_{b}+\delta)-g^{2},
\end{equation}
where $\delta\to0^{+}$ to ensure causality. A null eigenvalue can only
occur if $\omega_{b}=-\omega_{a}$, otherwise $\Lambda$ would
be complex. Since $\omega_{a}^{2}\geq0$ the equation 
\begin{equation}
(1+\delta)^{2}+\omega_{a}^{2}-g^{2}=0.
\end{equation}
does not have solution for $g<1$. The time-independent solution
$\Delta=0$ is hence stable for $g<1$, however it becomes unstable for
$g>1$.

For $g<1$ only the solution $\Delta=0$ exists. When $g>1$
a non trivial $\Delta > 0$ solution to Eq. \eqref{eq:t-ind-sol} exists.  The fluctuations
around this solution consists of two different branches.

The first are diagonal, \emph{within replica}, fluctuations 
$\Psi^{\bs{ab}}=\Psi(t_{a},t_{b})\,\delta^{\rm Kr}_{\alpha\beta}$.
The eigenvalue equation \eqref{eq:eig_eq_st} then becomes:
\begin{equation}
\begin{split}
\Bigl[(1&+\partial_{a})(1+\partial_{b})  -g^{2}[(\phi')^{2}]_{\Delta}\Bigr]\Psi(t_{a},t_{b})\\
 & -\frac{g^{2}}{2}[\phi\phi'']_{\Delta}\bigl[\Psi(t_{a},t_{a})+\Psi(t_{b},t_{b})\bigr]=\Lambda\Psi_{{\rm S}}(t_{a},t_{b}).
\end{split}
\end{equation}
The eigenfunctions are of the form $\Psi_{{\rm S}}(t_{a},t_{b})=\Psi(t_{a})+\Psi(t_{b})$.
Taking the Fourier transform with respect to time we find 
\begin{equation}
\Lambda=1+\delta-i\omega-g^{2}\Bigl[[(\phi')^{2}]_{\Delta}+[\phi\phi'']_{\Delta}\Bigr].
\end{equation}
For $\omega=0$ and $\phi(x)=\tanh(x)$ it is well known that $\Lambda > 0$ for all $g$.
In the theory of spin glasses this is equal to the second
eigenvalue of the Hessian of the fluctuations of the replica symmetric
solution of the Sherrington-Kirkpatrick (SK) model, see Ref. \cite{deAlTho78}. 
Thus the static solution is stable against \emph{within} replica
fluctuations. Note that this implies that in a one-replica system the time-independent 
solution is stable for all $g$.

Equation \eqref{eq:eig_eq_st} admits also off-diagonal, \emph{between replica}, fluctuations
$\Psi^{\bs{ab}}=\Psi(t_{a},t_{b})\,(1 - \delta^{\rm Kr}_{\alpha\beta})$.
For these fluctuations the eigenvalue equation \eqref{eq:eig_eq_st}
reduces to: 
\begin{equation}
\Bigl[(1+\partial_{a})(1+\partial_{b})-g^{2}[(\phi')^{2}]_{\Delta}\Bigr
         ]\Psi(t_{a},t_{b})=\Lambda\Psi(t_{a},t_{b}).
\end{equation}
Fourier transforming we find as before: 
\begin{equation}
\Lambda=(1-i\omega_a+\delta)(1-i\omega_b+\delta)-g^{2}[(\phi')^{2}]_{\Delta},
\end{equation}
which for $\omega_b=-\omega_a$ gives: 
\begin{equation}
\Lambda=(1+\delta)^{2}+\omega_a^{2}-g^{2}[(\phi')^{2}]_{\Delta}.
\end{equation}
The quantity $1-g^{2}[(\phi')^{2}]_{\Delta}$ with $\phi(x)=\tanh(x)$
appears also in the mean field theory of spin glasses. There it is
the relevant eigenvalue of the Hessian of the fluctuations of the
replica symmetric solution of the SK model , see Ref. \cite{deAlTho78}, 
and it is known to be negative for $g>1$. 
Thus $\Lambda$ can vanishes  for some $\omega_a$ and
the time-independent solution $\Delta^{\alpha\beta}=C^{\alpha\beta}=\Delta$
is unstable for $g>1$.

\subsection{\label{sec:Stab_tds}Stability of time-dependent solutions}

The stability analysis  of the steady state solutions
follows the same path as the time-independent solutions and it shall
not be repeated in details.

As for the time-independent case, the relevant self-consistent solution of the
DMFT equations is replica independent: 
$\Delta^{\bs{ab}}=\Delta^{ab}=\Delta(\tau)$, $\tau=t_{a}-t_{b}$, 
where  $\Delta(\tau)$  is solution of \eqref{eq:Newton_eq}. 

By denoting with $\Delta=\Delta(\tau)$ and $\Delta_{0}=\Delta(\tau=0)$
the derivatives occurring in the eigenvalue equation \eqref{eq:Eigen_Eq}
can be written as 
\begin{equation}
\frac{\partial}{\partial\Delta^{ab}}\bigl\langle S^{a}S^{b}\bigr\rangle_{0}=1+\partial_{\Delta}^{2}V(\Delta;\Delta_{0}),
\end{equation}
\begin{equation}
\frac{\partial}{\partial\Delta^{aa}}\bigl\langle S^{a}S^{b}\bigr\rangle_{0}=\frac{1}{2}\partial_{\Delta_{0}}\partial_{\Delta}V(\Delta;\Delta_{0}),
\end{equation}
where $V(\Delta;\Delta_{0})$ is the potential \eqref{eq:class_pot}
function of $\Delta$ and $\Delta_0$.
Details are in Appendix \ref{app:eq_class_pot}. The eigenvalue equation \eqref{eq:Eigen_Eq}
then reads: 
\begin{equation}
\begin{split}\Bigl[\partial_{a}+\partial_{b}&+\partial_{a}\partial_{b}-  \partial_{\Delta}^{2}V(\Delta;\Delta_{0})\Bigr]\,\Psi^{\bs{ab}}\\
 & -\frac{1}{2}\partial_{\Delta_{0}}\partial_{\Delta}V(\Delta;\Delta_{0})\bigl[\Psi^{\bs{aa}}+\Psi^{\bs{bb}}\bigr]=\Lambda\Psi^{\bs{ab}}.
\end{split}
\end{equation}
Since $\Lambda$ would be complex if $\Psi^{\alpha\beta}(t_{a},t_{b})$
does not depend on $\tau=t_{a}-t_{b}$ we will restrict to fluctuations
depending only on $\tau$. 
Hence, making explicit the time dependence of
$\Psi^{\bs{ab}}$, we have the equation: 
\begin{equation}
\begin{split}
   \Bigl[2\delta&-\partial_{\tau}^{2}- 
        \partial_{\Delta}^{2}V(\Delta;\Delta_{0})\Bigr]\,\Psi^{\alpha\beta}(\tau)\\
      & -\frac{1}{2}\partial_{\Delta_{0}}\partial_{\Delta}V(\Delta;\Delta_{0})\bigl[\Psi^{\alpha\alpha}(0)
       +\Psi^{\beta\beta}(0)\bigr]=\Lambda\Psi^{\alpha\beta}(\tau).
\end{split}
\label{eq:Stab_TD}
\end{equation}
The term $\delta^{2}$ has been neglected because $\delta\to0^{+}$.

Again the critical fluctuations
are  off-diagonal: 
$\Psi^{\alpha\beta}(\tau)=\Psi(\tau)\, (1-\delta^{\rm Kr}_{\alpha\beta})$.
The second term in Eq. \eqref{eq:Stab_TD}
then vanishes and, defining $\epsilon = \Lambda - 2\delta$,
the eigenvalue equation reduces to a one dimensional
time-independent Schr\"odinger equation in the variable $\tau$: 
\begin{equation}
       {\cal H}_{\tau}\Psi(\tau) := \Bigl[-\partial_{\tau}^{2}+V_{QM}(\tau)\Bigr]\,\Psi_{{\rm }}(\tau)
                                                    =\varepsilon\Psi(\tau),
\label{eq:Schr}
\end{equation}
with the quantum mechanical potential:
\begin{widetext}
\begin{equation}
\begin{split}
V_{{\rm QM}}(\tau) & =-\partial_{\Delta}^{2}V(\Delta;\Delta_{0})\Bigl|_{\Delta=\Delta(\tau)}\\
 & =1-g^{2}\int Dz\left[\int Dx\,\phi'\Bigl(gx\sqrt{\Delta_{0}-|\Delta|}+gz\sqrt{|\Delta|}\Bigr)\right]^{2}\Bigl|_{\Delta=\Delta(\tau)}
\end{split}
\label{eq:VQM}
\end{equation}
\end{widetext}
where $\Delta(\tau) \equiv \Delta(\tau;\Delta_0)$ is the solution to Eq. \eqref{eq:Newton_eq}.
\begin{figure}[tbh]
\centering \includegraphics[scale=0.6]{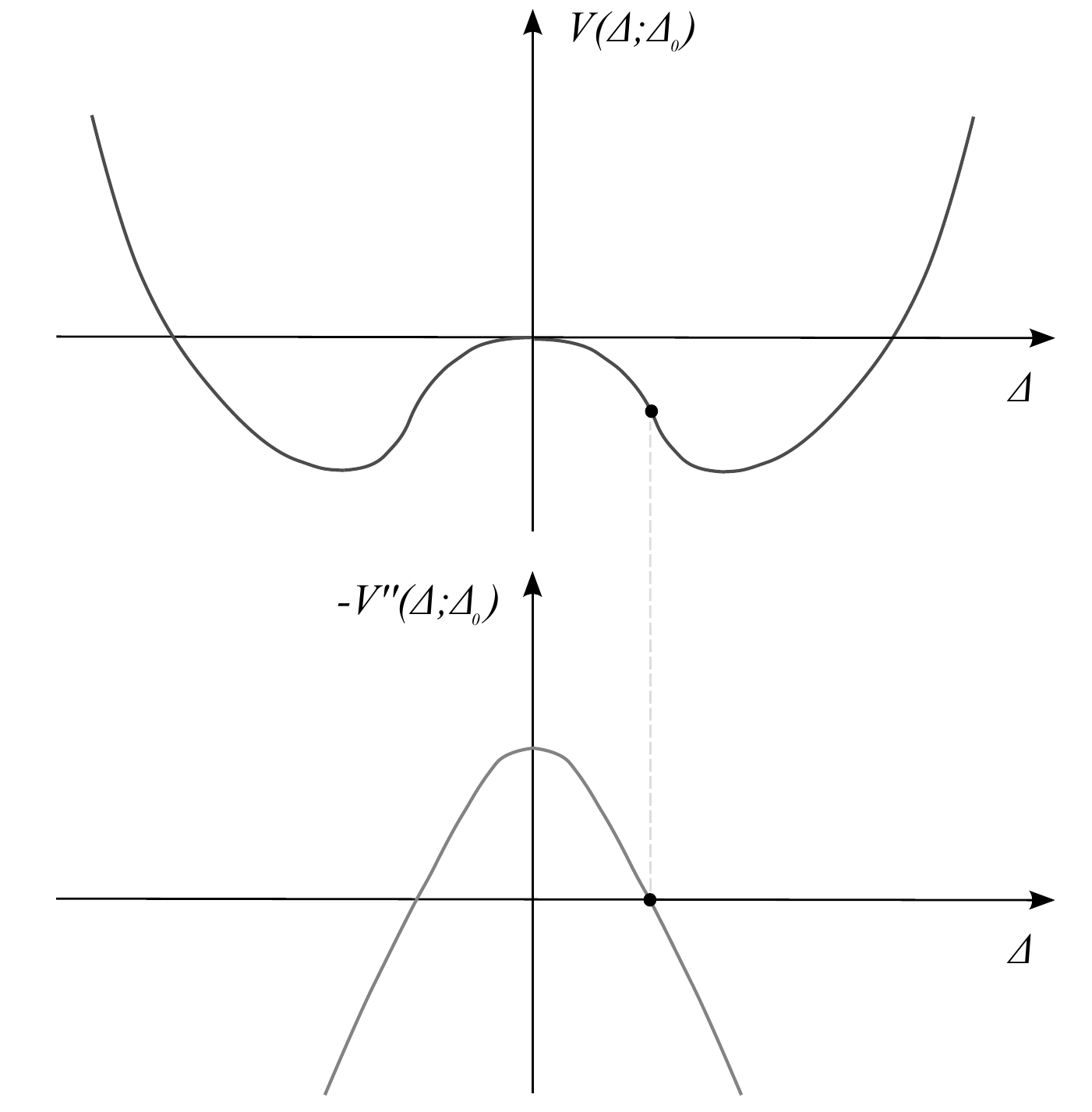} \caption{ Qualitative construction of the quantum potential \protect{\eqref{eq:VQM}}
from the potential \protect{\eqref{eq:class_pot}}. }
\label{fig:QMPot} 
\end{figure}

Equation \eqref{eq:Schr} always admits the eigenvalue $\varepsilon=0$ with the
eigenfunction $\Psi^{0}(\tau)=\partial_{\tau}\Delta(\tau)$, as
follows by differentiating Eq. \eqref{eq:Newton_eq} with respect to
$\tau$. This corresponds to eigenvalue $\Lambda=2\delta$, which is marginally
positive for $\delta\to0^{+}$.

The structure of the other eigenvalues depends on the form of the
quantum potential 
which ultimately depends on the solution 
$\Delta(\tau;\Delta_0)$, see Fig. \ref{fig:QMPot}.

\begin{figure}[tbh]
\centering \includegraphics[scale=0.65]{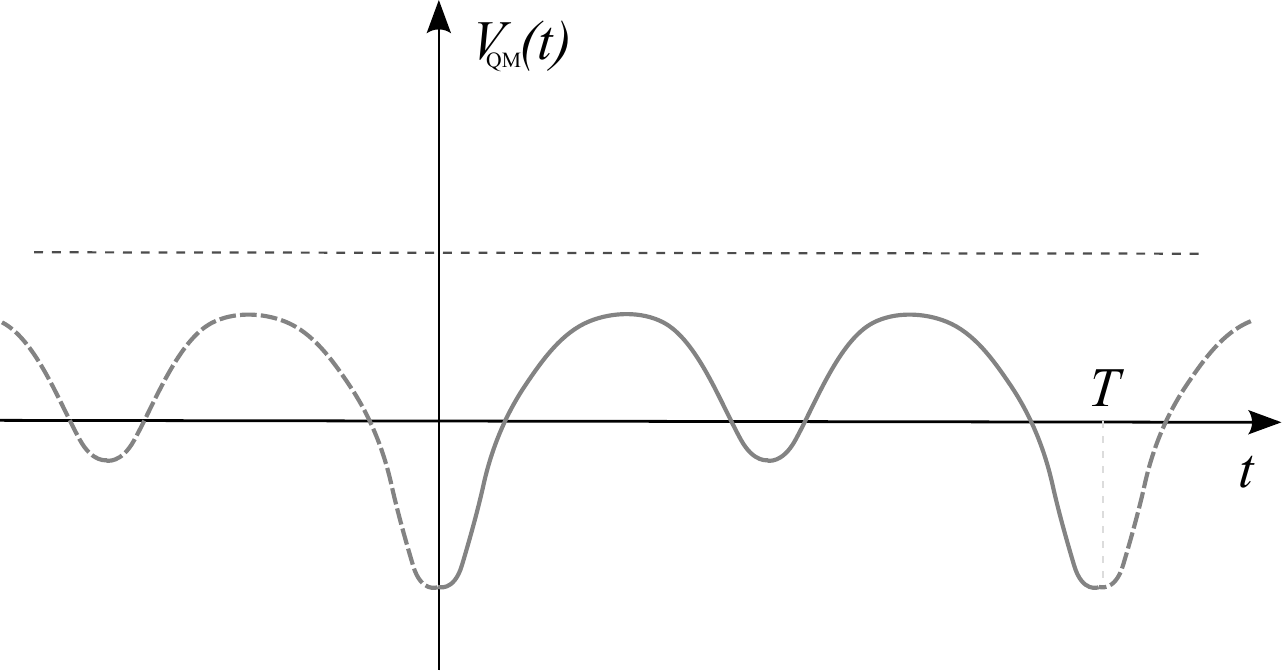} 
\caption{ Qualitative behaviour of the quantum potential \protect{\eqref{eq:VQM}}
for  time periodic solutions. }
\label{fig:QMPot-Per} 
\end{figure}

\textbf{Time Periodic solutions:} For time periodic solutions the quantum
potential $V_{{\rm QM}}(\tau)$ is periodic with the qualitative form
shown in Fig. \ref{fig:QMPot-Per}. The eigenfunction $\Psi^{0}(\tau)$
is also periodic and changes sign once within one period $T$, vanishing
at $\tau=0,T$. Thus there is \textit{exactly} one periodic eigenfunction
of ${\cal H}_{\tau}$ with eigenvalue $\varepsilon_{0}<0$ which vanishes
only at $\tau=0,T$. However since the potential is periodic there are 
bands of solutions where $\epsilon_{0}$ is the bottom of the
lowest band and $\varepsilon=0$ the top of the next band. Therefore
the eigenvalue $\Lambda=\varepsilon+2\delta$ passes continuously
through zero whatever small $\delta$ is, and hence the time periodic
solutions are unstable. 

\begin{figure}[tbh]
\centering \includegraphics[scale=0.65]{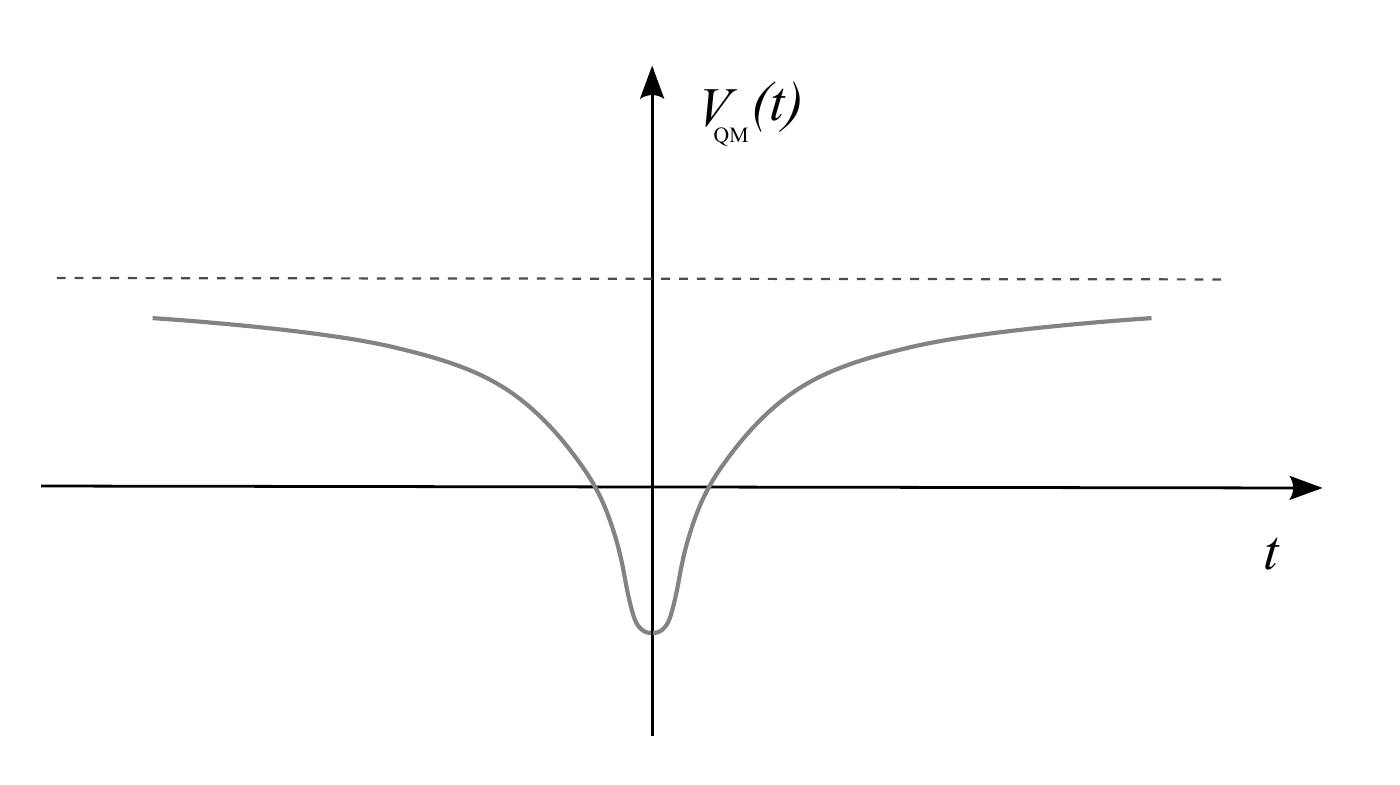} 
\caption{ Qualitative behaviour of the quantum potential \protect{\eqref{eq:VQM}}
for the time decaying solution. }
\label{fig:QMPot-Chaos} 
\end{figure}

\textbf{Time decaying solution:} If $\Delta(\tau)$ is the time decaying
solution $V_{{\rm QM}}(\tau)$ has the qualitative  form shown in Fig.
\ref{fig:QMPot-Chaos}. Again the eigenfunction
$\Psi^{0}(t)$ has exactly one node and from elementary quantum mechanics
we know that there is \textit{exactly} one eigenfunction of ${\cal H}_{\tau}$
with no nodes and eigenvalue $\varepsilon_{0}<0$. 
However in this case the eigenvalues of $V_{{\rm QM}}(\tau)$ are isolated and 
$\Lambda=\varepsilon_{0}+2\delta$, $\delta\to0^{+}$, cannot be zero
and the solution is stable. 

Summarising: when $g>1$ only the time decaying solution represents
a stable attractor of the dynamics.

We conclude this Section by noticing that 
the  Lagrangian \eqref{eq:S2} can be used also
to evaluate correlation functions of fluctuations around the mean
field as well as response functions. In the next Section we are using it to calculate
the Lyapunov exponent of the time decaying solution and prove that
it represents a chaotic state.

\section{\label{sec:Lyap}Maximum Lyapunov Exponent}

The decay of the time dependent $h$-correlation function $\Delta(\tau)$
suggests that the underlying neural dynamics is chaotic. A chaotic
dynamics exhibits an exponential sensitivity to initial conditions
A measure of the extent to which the dynamics is chaotic is provided
by the maximal Lyapunov exponent which measures the sensitivity of
the dynamics to small changes in the initial condition. To evaluate
this exponent, we  consider a small (infinitesimal) change in the state of the
system at time $t_{0}$ by $\delta h_{i}(t_{0})$, $i=1,...,N$. After
a time $t$ the perturbation has grown as 
\begin{equation}
\abs{\delta h(t)}\sim\abs{\delta h(t_{0})}\,e^{\lambda(t-t_{0})},
\end{equation}
where $\lambda$ is the maximal Lyapunov exponent. Positive $\lambda$
implies that the dynamic is chaotic.

As long as the perturbation is small, the perturbed trajectory $h_{i}(t)+\delta h_{i}(t)$
is close to the unperturbed trajectory, and the time evolution of
$\delta h_{i}(t)$ is ruled by the differential equation, 
\begin{equation}
\partial_{t}\delta h_{i}(t)=-\delta h_{i}(t)+g\sum_{j=1}^{N}J_{ij}\,\phi'\bigl(gh_{j}(t)\bigr)\,\delta h_{j}(t),
\end{equation}
obtained linearising the dynamical equations \eqref{eq:kirch} about
the unperturbed trajectory $h_{i}(t)$. The solution to this linear
equation with initial condition $\delta h_{i}(t_{0})$ can be written
as: 
\begin{equation}
\delta h_{i}(t)=\sum_{j=1}^{N}\chi_{ij}(t,t_{0})\,\delta h_{j}(t_{0}),
\end{equation}
where
\begin{equation}
\chi_{ij}(t,t')=\delta h_{i}(t)/\delta\bb_{j}(t'),
\qquad t\geq t',
\end{equation}
is the linear response of $h_i(t)$ to an infinitesimal perturbation
in the form of a small external field $\delta\bb_{j}(t')$ added to
 the rhs. of the dynamical equations \eqref{eq:kirch} at earlier time $t'$. 
 From
the multiplicative ergodic theorem of Oseledec the Lyapunov of the
stationary dynamics is \cite{LL83et}
\begin{equation}
   \lambda=\lim_{t-t_0\to\infty}\frac{1}{2(t-t_0)}\ln\left[\frac{1}{N}\sum_{ij}\chi_{ij}(t,t_{0})^{2}\right],
\end{equation}
and gives the dominant exponential growing rate of the linear response
as $t-t_{0}\gg1$.

For finite systems the dynamics depends on the couplings $J_{ij}$.
Hence for finite $N$ the exponent $\lambda$ is a random quantity.
However, in the limit $N\to\infty$ the dynamics converges to a non-random
behaviour, as described by the DMFT, $\sum_{ij}\chi_{ij}(t,t_{0})^{2}/N$
becomes self-averaging and $\lambda$ converges to the non-random
value: 
\begin{equation}
         \lambda=\lim_{t-t_0\to\infty}
             \frac{1}{2(t-t_0)}\ln\left[\frac{1}{N}\sum_{ij}\overline{\chi_{ij}(t,t_{0})^{2}}\right].
\label{eq:lyap}
\end{equation}

The maximum Lyapunov exponent can be computed using the DFT developed so 
far. However, to illustrate the power, and limitations, of the intuitive construction of self-consistent
equations for fluctuations for $N\gg 1$,  we first present the intuitive calculation of $\lambda$.
This uses some results discussed in the previous sections supplemented by some reasonable 
assumptions.
The systematic approach using the DFT will be present  next.

\subsection{Intuitive calculation of the Lyapunov exponent.}
The quantity $\sum_{ij}\overline{\chi_{ij}(t,t_0)^{2}}/N$ appearing in Eq. \eqref{eq:lyap}
can be computed from the Green function 
\begin{equation}
G(t_{a},t_{b},t_{c},t_{d})=\frac{1}{N}\sum_{i,j}^{N}
   \overline{\chi_{ij}(t_{a},t_{c})\chi_{ij}(t_{b},t_{d})},
\label{eq:def G}
\end{equation}
by taking  $t_a=t_b = t$ and $t_c=t_d=t_0$.
An equation for $G(t_{a},t_{b},t_{c},t_{d})$  can be constructed noticing that
from the dynamical equation \eqref{eq:kirch} it follows that 
the linear response $\chi_{ij}(t,t')$ obeys the differential equation:
\begin{equation}
\begin{split}
     \left(1+\frac{d}{dt}\right)\chi_{ij}(t,t')=
          g\sum_{k=1}^{N}J_{ik}&\phi'\bigl(gh_{k}(t)\bigr)\, \chi_{kj}(t,t')
          \\
          & +\delta(t-t')\,\delta^{\rm Kr}_{ij},
\end{split}          
\label{eq:chi_dynamics}
\end{equation}
where the  Kronecker and Dirac
delta functions denote a local spatiotemporal perturbation.
Thus, by multiplying Eq. \eqref{eq:chi_dynamics} by itself and taking the spatial average, we find:
\begin{multline}
\left(1+\frac{\partial}{\partial t_{a}}\right)\left(1+\frac{\partial}{\partial t_{b}}\right)G\left(t_{a},t_{b},t_{c},t_{d}\right) \\
-\frac{\partial C\left(t_{a}-t_{b}\right)}{\partial\Delta\left(t_{a}-t_{b}\right)}G\left(t_{a},t_{b},t_{c},t_{d}\right) \\ 
= \delta\left(t_{a}-t_{b}-t_{c}+t_{d}\right)\delta\left(t_{a}+t_{b}-t_{c}-t_{d}\right).\label{eq:G_abcd_dynamics}
\end{multline}
To arrive at this equation we have used the fact that under averaging 
$\phi'(t)\phi'(t')$ can be replaced by $\partial C\left(t-t'\right) / \partial\Delta\left(t-t'\right)$
and assumed that the cross term in squaring Eq. \eqref{eq:chi_dynamics}
vanishes  or, equivalently, vanishes in the
large $N$ limit due to the summation in Eq. \eqref{eq:def G}. 

Defining the new time variables
\begin{equation}
s=t_{a}+t_{b},\quad s'=t_{c}+t_{d},\label{eq:times}
\end{equation}
\begin{equation}
\tau=t_{a}-t_{b},\quad\tau'=t_{c}-t_{d}.\label{eq:times2}
\end{equation}
equation \eqref{eq:G_abcd_dynamics} can be written as 
\begin{equation}
 \Bigl[ 2\partial_s + \partial_s^2  + \mathcal{H}_{\tau}\Bigr]\,G(s,s',\tau,\tau')
           = 2\,\delta(s-s')\,\delta(\tau-\tau'),
\label{eq:G_dynamics}
\end{equation}
where ${\cal H}_{\tau}=-\partial_{\tau}^{2}-\partial_{\Delta}^{2}V(\Delta;\Delta_{0})$
is the quantum mechanical Hamiltonian acting on variable $\tau$
introduced in Sec. \ref{sec:Stab_tds}. 
The solution to this equation can be written as
\begin{equation}
G(s,s',\tau,\tau')=2\sum_{n}g_{n}(s,s')\,\varphi_{n}(\tau)\,\varphi_{n}^{*}(\tau'),
 \label{eq:PhQ-1}
\end{equation}
where $\varphi_n(\tau)$ is the set of orthonormal eigenfunctions  of ${\cal H}_\tau$:
${\cal H}_\tau\, \varphi_n = \epsilon_n\, \varphi_n$, and the 
function $g(s,s')$  is solution of the differential equation:
\begin{equation}
\label{eq:gn}
   \Bigl[2\partial_{s}+\partial_{s}^{2}+\varepsilon_{n}\Bigr]\,g_{n}(s,s')=\delta(s-s').
\end{equation}
The sum in Eq. \eqref{eq:PhQ-1} is over all eigenfunctions of $\mathcal{H}_\tau$
and may include a continuum part of the spectrum of ${\cal H}_\tau$. 
Taking now $t_a=t_b = t$ and $t_c=t_d=t_0$ we finally arrive at
\begin{equation}
     \frac{1}{N}\sum_{ij}\overline{\chi_{ij}(t,t_{0})^{2}} = 
        2\sum_{n}\, g_{n}(2t,2t_0)\,\varphi_{n}(0)\,\varphi_{n}^{*}(0).
\label{eq:chi2}
\end{equation}
The maximum Lyapunov exponent is related to the asymptotic behaviour of 
$g_n(t,t')$ as $t-t'\to \infty$.
While this equation leads to the correct $\lambda$, its derivation is clearly not systematic.
It is difficult to have control on the approximations and, moreover, in more complex 
cases it can be difficult to be constructed.
Therefore, before discussing the Lyapunov exponent, we show 
how equations like  \eqref{eq:G_dynamics} can 
systematically derived within the  DFT.

\subsection{DFT calculation of the Lyapunov exponent.}

Within the replica formalism introduced in Sec. \ref{sec:RDMFT},
it is more convenient to calculate Lyapunov exponent  via the spin
susceptibility,
\begin{equation}
  \tilde{\chi}_{ij}(t,t_{0}) =\left.\frac{\delta S_{i}(t)}{\delta\bb_{j}(t_{0})}\right|_{\bb=0}
                              =g\phi'(gh_{i})\,\chi_{ij}(t,t_{0}),
\label{eq:chitilde}
\end{equation}
the linear response of $S_i(t)$ to an infinitesimal external field $\delta\bb_{j}(t_0)$ 
at the earlier time $t_0$. 

The Lyapunov exponent $\lambda$ is related to the fluctuations 
$\sum_{ij}\overline{\tilde{\chi}_{ij}(t,t_{0})^{2}}/N$
of the spin susceptibility. 
Introducing two identical replicas of the system these can be computed in 
using the DFT as:
\begin{widetext}
\begin{equation}
\begin{split}
        \frac{1}{N}\sum_{ij}\overline{\left(
            \left.\frac{\delta\langle S_{i}(t)\rangle_{J\bb}}
                          {\delta\bb_{j}(t_{0})}\right|_{\bb=0}\right)^{2}} &
     =  \frac{1}{N}\sum_{ij}\overline{\bigl\langle S_{i}(t)i\hat{h}_{j}(t_{0})\bigr\rangle_{J}\,
                                                      \bigl\langle S_{i}(t)i\hat{h}_{j}(t_{0})\bigr\rangle}_{J}\\
 & =\frac{1}{N}\sum_{ij}\bigl\langle S_{i}^{\bs{a}}S_{i}^{\bs{b}}i\hat{h}_{j}^{\bs{c}}i\hat{h}_{j}^{\bs{d}}
                                     \bigr\rangle,
\end{split}
\end{equation}
\end{widetext}
with replica indexes $\alpha=\gamma\not=\beta=\delta$, 
$\alpha$ and $\beta$ being the replica indices of $\bs{a}$ and $\bs{b}$
while $\gamma$ and $\delta$ those of $\bs{c}$ and $\bs{d}$,
and time arguments $t_{a}=t_{b}=t$ and $t_{c}=t_{d}=t_{0}$. 

Evaluating the average leads to 
\begin{equation}
       \frac{1}{N}\sum_{ij} \bigl\langle S_{i}^{\bs{a}}S_{i}^{\bs{b}}i\hat{h}_{j}^{\bs{c}}i\hat{h}_{j}^{\bs{d}}
                                       \bigr\rangle
       = N \bigl\langle C^{\bs{ab}}i\hat{C}^{\bs{cd}}\bigr\rangle-\frac{1}{2}\,\delta_{\bs{ab},\bs{cd}}.
\label{eq:lyap2}
\end{equation}
Details can be found in Appendix \ref{app:lyap2}.
%
This relation is an exact relation valid for any $N$. In the limit $N\to\infty$ the two-point 
correlation function $\langle C^{\bs{ab}}i\hat{C}^{\bs{cd}}\rangle$ reduces to
the two-point correlation function $\langle Q^{\bs{ab}}i\hat{Q}^{\bs{cd}}\rangle$
of the Gaussian fluctuations about the saddle point. 
This can be evaluated  from the quadratic action 
${\cal L}_{2}[\hat{Q},\Psi;\hat{\bb},\bb]$
as 
\begin{equation}
     (1+\partial_{a})(1+\partial_{b})\bigl\langle\Psi^{\bs{ab}}i\hat{Q}^{\bs{cd}}\bigr\rangle
             =\bigl\langle Q^{\bs{ab}}i\hat{Q}^{\bs{cd}}\bigr\rangle,
\label{eq:PsiQ}
\end{equation}
where, from Eq. \eqref{eq:S2}, $\langle\Psi^{\bs{ab}}i\hat{Q}^{\bs{cd}}\rangle$
satisfies the equation 
\begin{equation}
      \sum_{\bs{ef}}\,{\cal A}^{\bs{ab};\bs{ef}}\bigl\langle\Psi^{\bs{ef}}i\hat{Q}^{\bs{cd}}\bigr\rangle
      = \frac{1}{2N}\delta_{\bs{ab}\bs{,cd}},
\label{eq:HPsiQ}
\end{equation}
with the operator ${\cal A}$ defined in Eq. \eqref{eq:ep_H}.

For the particular choice of replica indexes $\gamma=\alpha$ and $\delta=\beta$ but $
\alpha\not=\beta$,  and changing time variables as in Eqs. \eqref{eq:times} and \eqref{eq:times2},
equation \eqref{eq:PsiQ} becomes:
\begin{equation}
\label{eq:eq_PsiQ}
\begin{split}
  \Bigl[1 + 2\partial_s + \partial_s^2 - \partial_\tau^2\Bigr]&
      \bigl\langle\Psi^{\alpha\beta}(s,\tau)\,i\hat{Q}^{\alpha\beta}(s',\tau')\bigr\rangle \\
      & =
      \bigl\langle Q^{\alpha\beta}(s,\tau)\,i\hat{Q}^{\alpha\beta}(s',\tau')\bigr\rangle.
\end{split}      
\end{equation}
Similarly, working out the explicit form of ${\cal A}$ as done in Sec. \ref{sec:Stab_tds}, 
equation \eqref{eq:HPsiQ} becomes:
\begin{equation}
\begin{split}
    \Bigl[2\partial_{s}+\partial_{s}^{2}+{\cal H}_{\tau}\Bigr] &
        \bigl\langle\Psi^{\alpha\beta}(s,\tau)\,i\hat{Q}^{\alpha\beta}(s',\tau')\bigr\rangle
           \\
           & =\frac{1}{N}\delta(s-s')\,\delta(\tau-\tau'),
\end{split}            
\label{eq:eq_HPsiQ}
\end{equation}
with the quantum mechanical Hamiltonian ${\cal H}_{\tau}$
acting on  the (time) variable $\tau$. 

The solution to Eq. \eqref{eq:eq_HPsiQ} reads:
\begin{equation}
    \bigl\langle\Psi^{\alpha\beta}(s,\tau)\,i\hat{Q}^{\alpha\beta}(s',\tau')\bigr\rangle
    =\frac{1}{N}\sum_{n}g_{n}(s,s')\,\varphi_{n}(\tau)\,\varphi_{n}^{*}(\tau'),
\end{equation}
where $\varphi_{n}(\tau)$ are the orthonormal eigenfunctions of ${\cal H}_\tau$ and 
$g_n(s,s')$ is the solution of the differential equation \eqref{eq:gn}.
On the other hand, 
subtracting Eq. \eqref{eq:eq_HPsiQ} from Eq. \eqref{eq:eq_PsiQ} leads to:
\begin{widetext}
\begin{equation}
  \bigl\langle Q^{\alpha\beta}(s,\tau)\,i\hat{Q}^{\alpha\beta}(s',\tau')\bigr\rangle
  = 
  \Bigl[1 
  -V_{\rm QM}(\tau)
  \Bigr]
   \bigl\langle\Psi^{\alpha\beta}(s,\tau)\,i\hat{Q}^{\alpha\beta}(s',\tau')\bigr\rangle
   + \frac{1}{N}\delta(s-s')\,\delta(\tau-\tau').
\end{equation}
\end{widetext}
Inserting these expressions into Eq. \eqref{eq:lyap2} gives:
\begin{equation}
\begin{split}
\frac{1}{N}\sum_{ij} & \bigl\langle S_{i}^{\alpha}(t_{a})S_{i}^{\beta}(t_{b})
     i\hat{h}_{j}^{\alpha}(t_{c})i\hat{h}_{j}^{\beta}(t_{d})\bigr\rangle
     \\
     &=
     \Bigl[1 - V_{{\rm QM}}(\tau)\Bigl]\sum_{n}g_{n}(s,s')\,\varphi_{n}(\tau)\,\varphi_{n}^{*}(\tau'),
\end{split}     
\end{equation}
and taking $t_{a}=t_{b}=t$ and $t_{c}=t_{d}=t_{0}$ finally leads to:
\begin{equation}
\frac{1}{N}\sum_{ij}\overline{\tilde{\chi}_{ij}(t,t_{0})^{2}}=
    \Bigl[1 - V_{{\rm QM}}(0)\Bigl]\sum_{n}g_{n}(2t,2t_{0})\,\varphi_{n}(0)\,\varphi_{n}^{*}(0),
\end{equation}
which is identical to Eq. \eqref{eq:chi2} apart from a constant, related
to the transformation \eqref{eq:chitilde} between the two susceptibilities.

The solution to Eq. \eqref{eq:gn} which vanishes for $s<s'$ is: 
\begin{equation}
       g_{n}(s,s')=\frac{\theta(s-s')}{\sqrt{1-\varepsilon_{n}}}\,e^{-(s-s')}
       \sinh\Bigl[\sqrt{(1-\varepsilon_{n})}(s-s')\Bigr].
\end{equation}
Thus in the limit $t-t_{0}\gg1$: 
\begin{equation}
\frac{1}{N}\sum_{ij}\overline{\tilde{\chi}_{ij}(t,t_{0})^{2}}\sim
\Bigl[1 - V_{{\rm QM}}(0)\Bigl]\sum_{n}\frac{\varphi_{n}(0)\,\varphi_{n}^{*}(0)}{\sqrt{1-\epsilon_{n}}}\,e^{2\lambda_{n}(t-t_{0})},
\end{equation}
with $\lambda_{n}=-1+\sqrt{1-\epsilon_{n}}$, and hence the maximal
Lyapunov exponent is 
\begin{equation}
\lambda=\max_{n}\lambda_{n}=-1+\sqrt{1-\epsilon_{0}},
\end{equation}
where $\epsilon_{0}$ is the lowest eigenvalue of ${\cal H}_{\tau}$.

For $g<1$ we have seen that only the time-independent solution $\Delta=0$
exists. In this case $V_{{\rm QM}}=1-g^{2}$, see Eq. \eqref{eq:VQM},
therefore $\epsilon_{0}=1-g^{2}$ and $\lambda=-1+g<0$, showing that
$\Delta=0$ is a stable fix point for $g<1$. When $g>1$ the stable
solution is the time-dependent decaying solution 
which leads to a negative $\epsilon_{0}$.
The Lyapunov exponent is then positive and
the solution is \textit{chaotic}.

\section{Explicit expressions for $\lambda$ in time dependent state }

To find the expression of $\lambda$  we first have to solve the
DMFT equation \eqref{eq:Newton_eq} and then find the lowest 
eigenvalue of the associated quantum mechanical problem:
\be
         {\cal H}_{\tau}\Psi(\tau) := \Bigl[-\partial_{\tau}^{2} 
            - \partial_\Delta^2 V(\Delta;\Delta_0)\Bigr]\,\Psi_{{\rm }}(\tau)
                                                    =\varepsilon\Psi(\tau).
\ee
This is not an easy task for an arbitrary $g>1$.
However in the limit $g\to1^{+}$ and $g\to\infty$
the leading behaviour of $\lambda(g)$ can be determined, as shown below.

\subsection{Limit $g\to1^{+}$.}
The energy \eqref{eq:Ec} of the decaying DMFT is $E_c=0$.
The solution to the DMFT equation \eqref{eq:Newton_eq} can then be written in the
implicit form as:
\begin{equation}
      \tau =-\int_{\Delta_{0}}^{\Delta}\frac{d\Delta}{\sqrt{-2V(\Delta;\Delta_{0})}}.
\label{eq:D_dec}      
\end{equation}
In  the limit $g\to1^{+}$ the equal-time field correlation  $\Delta_0$ vanishes, thus
$\abs{\Delta}\leq \Delta_0 \ll1$ for all $t$ as $\sigma = g - 1 \ll 1$. 
Expanding the potential $V(\Delta;\Delta_{0})$ in powers of $\Delta$ and $\Delta_{0}$
to the leading non-trivial (fourth) order gives:
\begin{equation}
V(\Delta;\Delta_{0})\sim\Bigl(-1+g^{2}-2g^{4}\Delta_{0}+5g^{6}\Delta_{0}^{2}\Bigr)\frac{\Delta^{2}}{2}+g^{6}\frac{\Delta^{4}}{6}.
\end{equation}
The value of $\Delta_{0}$ is  found from the condition $V(\Delta_{0};\Delta_{0})=0$,
and reads $\Delta_{0}\sim\sigma-4\sigma^{2}/3+O(\sigma^{3})$ as 
$\sigma \to 0^+$. Thus 
\begin{equation}
V(\Delta;\Delta_{0})\sim-\frac{\sigma^{2}}{6}\Delta^{2}+\frac{1}{6}\Delta^{4},\qquad\sigma\to0^{+}.
\end{equation}
Substituting this expression into Eq. \eqref{eq:D_dec} leads to:
\begin{equation}
    \tau \sim\frac{\sqrt{3}}{\sigma}\int_{\Delta_{0}/\sigma}^{\Delta/\sigma}\frac{dx}{x\sqrt{1-x^{2}}},
    \qquad\sigma\to0^{+},
\end{equation}
which to the leading term in $\sigma$ gives: 
\begin{equation}
\Delta(\tau)=\sigma\cosh^{-1}\left(\frac{\sigma \tau}{\sqrt{3}}\right)+O\Bigl(\sigma^{3/2}\Bigr),
    \qquad\sigma\to0^{+}.\label{eq:Dg1+}
\end{equation}
Note that as $\sigma\to0^{+}$ the amplitude of $\Delta(\tau)$ vanishes linearly with $\sigma$ while
the characteristic decaying time diverges as $\sigma^{-1}$.
Thus as $g\to 1^+$ the dynamics slows down and the chaotic
attractor goes continuously to the fix point $\Delta=0$ at the critical point $g=1$.

Evaluating the quantum potential $V_{\rm QM}(\tau) = -\partial_{\Delta}^{2}V(\Delta;\Delta_{0})|_{\Delta=\Delta(\tau)}$
relative to the solution \eqref{eq:Dg1+} the associated quantum mechanical problem
becomes: 
\begin{equation}
\left[\partial_{\tau}^{2}
       -2\sigma^{2}\left[\cosh^{-1}\left(\frac{\sigma \tau}{\sqrt{3}}\right)\right]^{2}\right]\varphi_{n}= 
         \left(\epsilon_{n}-\frac{\sigma^{2}}{3}\right)\varphi_{n}.
\end{equation}
The solution to this differential equation are the generalised Legendre
functions with eigenvalues $\epsilon_{n}=-\sigma^{2}\bigl[(2-n)^{2}-1\bigr]/3$,
see e.g. Ref. \cite{LL}. Thus $\epsilon_{0}=-\sigma^{2}$ and 
\begin{equation}
\lambda=-1+\sqrt{1-\epsilon_{0}}\sim\frac{1}{2}(g-1)^{2},
    \qquad g \to 1^{+}.
\end{equation}
Notice that near the onset of chaos the rate $\lambda^{-1}$ of the exponential 
divergence of close-by  trajectories scales as the square of the rates of the 
decay of memory along the chaotic trajectory.

\subsection{Limit $g\to\infty$}
The quantum potential behaves as $V_{{\rm QM}}(\tau)\sim-g$ for $\tau=O(1/g)$,
while  it converges  to a finite values independent of the value of $g$
as $\tau\to \pm\infty$.
Thus in the limit $g\gg1$
the potential  $V_{{\rm QM}}(\tau)$ becomes a very deep and narrow potential well close
to $\tau=0$, see Fig. \ref{fig:QMPot-Chaos}. The ground state eigenfunction
$\varphi_{0}(\tau)$ is localised in a region of width $O(1/g)\ll1$
at $\tau=0$ and decays exponentially fast outside this region.

In this scenario the leading behavior of the lowest eigenvalue $\epsilon_{0}$
of ${\cal H}_{t}$ as $g\to\infty$ can be obtained replacing the
original quantum mechanical problem by 
\begin{equation}
\Bigl[-\partial_{\tau}^{2}-V_{0}\,\delta(t)\Bigr]\, \varphi_{0}(\tau)=\epsilon_{0}\,\varphi_{0}(\tau),
\end{equation}
where 
\begin{equation}
    -V_{0}=\int_{-\Lambda}^{+\Lambda}d\tau\,V_{{\rm QM}}(\tau)
              = 2\int_{0}^{+\Lambda}d\tau\,V_{{\rm QM}}(\tau),
\end{equation}
because the quantum potential is an even function of $\tau$.
The parameter $\Lambda=O(1)$ is an arbitrary cut-off whose precise value is
irrelevant as long as we are interested in the leading behavior
as $g\gg1$. The solution to this equation is $\varphi_{0}(\tau)\propto\exp(-\sqrt{-\epsilon_{0}}|\tau|)$,
with $\epsilon_{0}=-(V_{0}/2)^{2}$.

To compute $V_{0}$ we introduce a point $a/g$, where $a$ is an arbitrary positive
constant, and split the integration as 
\begin{equation}
   -\frac{V_{0}}{2}=\int_{0}^{a/g}d\tau\,V_{{\rm QM}}(\tau) 
                    + \int_{a/g}^{+\Lambda}d\tau\,V_{{\rm QM}}(\tau).
\label{eq:V0}                    
\end{equation}
The first integral is $O(1)$ as $g\gg 1$ because $V_{\rm QM}(\tau) = O(g)$ is this region.
Thus the leading behaviour of $V_0$ as $g\gg 1$ is fully determined by the behaviour of 
$V_{\rm QM}(\tau)$ as $\tau = O(1/g)$.

Expanding $\Delta(\tau)$ about $\tau=0$ we find to the leading order in $\tau$: 
\begin{equation}
   \Delta(\tau) = \Delta_{0} + (\Delta_{0}-1)\,\frac{\tau^{2}}{2}+O(\tau^{3}),
   \qquad \tau \ll 1.
\label{eq:Delta_small_t}
\end{equation}
To obtain this expression we have used the initial condition $\partial_{\tau}\Delta(\tau)|_{\tau=0}=0$,
the DMFT equation \eqref{eq:Newton_eq} to evaluate $\partial_{\tau}^2\Delta(\tau)|_{\tau=0}$ and
\be
\begin{split}
   \partial_\Delta V(\Delta;\Delta_{0})|_{\Delta =\Delta_0} & = -\Delta_0 + [\phi^2]_{\Delta_0} 
   \\
   & \sim -\Delta_0 + 1 + O(1/g),
   \qquad g \gg 1.
\end{split}   
\ee
The value of $\Delta_0$ is again fixed by the requirement $V(\Delta_{0};\Delta_{0})=0$,
which as $g\to\infty$ gives $\Delta_{0}=2(1-2/\pi)$ to the leading order.

Using Eq. \eqref{eq:Delta_small_t} leads the following 
asymptotic expansion of $V_{\rm QM}(\tau)$ 
valid for $g\gg1$ and $\tau$ to $O(1/g)$: 
\begin{equation}
    V_{{\rm QM}}(\tau)\sim-\frac{C}{\tau}+1+O(1/g^{2}),
\end{equation}
where $C=\frac{2}{\pi}/\sqrt{\Delta_{0}(1-\Delta_{0})}$.
Thus from Eq. \eqref{eq:V0} it follows,
\begin{equation}
   -\frac{V_{0}}{2} \sim C\ln q+O(1),
  \qquad g\gg1,
\end{equation}
so that:
\begin{equation}
     \lambda=-1+\sqrt{1-\epsilon_{0}}\sim C\, \ln g,
     \qquad g\gg1.
\end{equation}

Notice that while the rate of  exponential divergence of close-by trajectories 
vanishes in the large $g$ limit,  the decay rate of memory along a trajectory remains finite.
Indeed, in the limit  $g\to\infty$ the DMFT equation \eqref{eq:Newton_eq} becomes 
\begin{equation}
\partial_{\tau}^{2}\Delta=\Delta-\frac{2}{\pi}\sin^{-1}\left(\frac{\Delta}{\Delta_{0}}\right)
                                \underset{\Delta\ll1}{\sim} \left(1 - \frac{2}{\pi\Delta_0}\right)\, \Delta,
\end{equation}
so that $\Delta(\tau)$ decay exponentially for $\tau\gg1$ with a characteristic time 
$\sqrt{1 - 2/\pi\Delta_0}$, cf. Ref. \cite{CriSom88}. 

\section{Discussion}

In this paper we have described a systematic approach to the dynamics of randomly connected 
neural networks based on the Path Integral Formalism originally introduced to study the stochastic 
dynamics in statistical mechanics. The problem of studying the dynamical behavior of the networks 
is formulated in terms of a dynamical field theory. For the sake of simplicity, we focused on a class 
of network models with simple architecture and odd-symmetric sigmoidal nonlinearity, as in model 
\eqref{eq:kirch} and \eqref{eq:coup_mom}. Using the Path Integral formalism, we have shown how 
the DMF equations can be derived as a saddle point of the path integrals, which becomes exact in 
the large N limit. Next, we studied the fluctuations around the saddle point and derived expressions 
for the multiple response and correlation functions. This fluctuation analysis yielded stability 
conditions for the stability of the DMF solutions. Finally, using the well-known relations between the 
maximal Lyapunov Exponent of a dynamical system to an appropriate linear response function, we 
derived equations for the Lyapunov exponent of the random network. Interestingly, in this simple 
network, the DMF equations for the order parameter bear a mechanical analog of a conservative 
Newtonian dynamics, whereas the susceptibility associated with the Lyapunov exponent have a 
quantum mechanical analog in the form of one dimensional (which is time) Schrodinger equation. 
In the simplest network architectures and dynamics, such as model 
\eqref{eq:kirch}-\eqref{eq:coup_mom}, the DMF equations can be derived by an intuitive
construction of the self-consistent equation governing the fluctuations in the system, using 
gaussianity ansatz of the fluctuating synaptic fields. Likewise, heuristic assumptions about the 
statistics of response functions can be used to calculate the maximal Lyapunov exponent, as we 
have shown here. However, this heuristic approach suffers from considerable limitations. First, it is 
hard to control the underlying ad-hoc assumptions. Notably, the extension to more complex 
connectivity or dynamics may be difficult to derive by heuristic methods, as for example the case of 
connections which are not fully asymmetric, or dynamics involving non-gaussian stochasticity (e.g., 
Poisson neurons). An additional difficulty is deriving stability conditions for the DMF solutions. As 
shown in Ref. \cite{KadSom15}, even the derivation of stability conditions for fixed points in random 
networks with more complex architecture may be quite challenging. 
Finally, in principle, the path integral method can be used to study systematic perturbations analysis 
to a finite-dimensional systems as well as systematic finite size corrections. Such applications of the 
path integral methods have been extensively developed for non-random stochastic dynamics in 
statistical mechanics, as well as in spin glasses. It will be very interesting to explore these directions 
in deterministic dynamics of random neural networks.

\begin{acknowledgments}
This work originates from a collaboration with H.-J. Sommers which led the joint publication in Ref. 
\cite{SomCriSom88}.   
The work of HS is partially supported by the Gatsby Charitable Foundation and the NIH. 
\end{acknowledgments}

\appendix
\section{Chaotic behaviour in the Ising limit}
\label{app:Amari}

In the Ising limit $g\to\infty$ the spin (or field) autocorrelation function decays exponentially in time 
with a finite characteristic time $\tau_a$ \cite{CriSom88}. Thus if we discretize the time in step
$\delta t\sim\tau_a$
the
evolution of the model is described by the $N$-dimensional map,
\begin{equation}
     h_i(n+1)=(1-\delta t)\,h_i(n)+
              \delta t\sum_{j=1}^{N}\ J_{ij} S\bigl(h_j(n)\bigr).
\end{equation}
The maximal Lyapunov exponent $\lambda$ is obtained from the
time-evolution of the tangent vector \cite{LL83et,ER85,BGGS80},
\begin{equation}
     \bm{\xi}(n+1)={\cal A}(n)\bm{\xi}(n),
\end{equation}
\begin{equation}
     {\cal A}_{ij}(n)=(1-\delta t)\,\delta_{ij}+
              g \delta t J_{ij}\ \cosh^{-2}(gh_j(n)).
 \label{eq:mod10}
\end{equation}
Since $\delta t\sim\tau_a$, we can assume that $h_i(n)$ and
$h_i(n')$ are uncorrelated if $n\not=n'$. 
Moreover, if $g\gg 1$ the leading contribution to ${\cal A}(n)$ comes from $|h_j|<1/g$.
Thus we can replace in Eq. \eqref{eq:mod10} $\cosh^{-2}$ by a constant, 
so that $\bm{\xi}(n)$ is given
by a product of independent $N \times N$ random matrices. In the limit
$N\gg 1$ the diagonal part of ${\cal A}(n)$ does not contribute and one has
\cite{CN84,CPV88},
\begin{equation}
   \lambda \sim \left\{\begin{array}{ll}
              \ln{ \overline{A_{ij}    } }, &\mbox{if $\overline{ A_{ij} }\not=0;$} \\
                   \\
              \ln{ \overline{ A_{ij}^2 } }, &\mbox{if $\overline{ A_{ij} }=0;$}
                       \end{array}
                 \right.
\end{equation}
where $\overline{(\cdot)}$ means averaging over the different realizations of
$J_{ij}$'s. Therefore,
\begin{equation}
    \lambda\sim \ln{g}, \qquad g\gg 1,
\end{equation}
and the dynamics is chaotic.

\section{ $\langle SS\rangle$ correlation function}
\label{app:eq_C_ab} 
The average $\langle{\cal F}[h^{a}]\rangle_{h}$
of any functional of $h_{i}^{a}$ over the solutions of the dynamical
equation \eqref{eq:eq_dyn} can be written as a path integral over
all trajectories $\{\hat{h}_{i},h_{i}\}_{t\in[t_{0},t]}$ weighted with the
dynamical action $S[\hat{h},h]$ \eqref{eq:msr_act}.
Thus:
\begin{equation}
\begin{split}
\sum_{i=1}^{N}\bigl\langle S_{i}(t_{a})S_{i}(t_{b})\bigr\rangle_{J}
     =\int\prod_{i}&{\cal D}\hat{h}_{i}\,{\cal D}h_{i}\,  e^{-S[\hat{h},h]}\\
     &\times 
     \sum_{i=1}^{N}S_{i}(t_{a})S_{i}(t_{b}),
\end{split}     
\end{equation}
Averaging over
the couplings $J_{ij}$, and introducing the auxiliary fields $C^{ab}$
and $\hat{C}^{ab}$, a straightforward calculation leads to: 
\begin{equation}
\begin{split}\sum_{i=1}^{N}&\overline{\bigl\langle S_{i}(t_{a})S_{i}(t_{b})\bigr\rangle_{J}}=\int  {\cal D}\hat{C}\,{\cal D}C\,e^{-N\sum_{(ab)}i\hat{C}^{ab}C^{ab}}\\
 & \times\int\prod_{i}{\cal D}\hat{h}_{i}\,{\cal D}h_{i}\,e^{-S[\hat{h}_{i},h_{i};\hat{C},C]}\sum_{i=1}^{N}S_{i}(t_{a})S_{i}(t_{b}),
\end{split}
\label{eq:AppB_ss_av_1}
\end{equation}
where $S[\hat{h}_{i},h_{i};\hat{C},C]$ is defined in Eq. \eqref{eq:DFT_hh_S}.
Using the identity 
\begin{equation}
         e^{-\sum_{i}S[\hat{h}_{i},h_{i};\hat{C},C]}\sum_{i=1}^{N}S_{i}(t_{a})S_{i}(t_{b})=
          \frac{\delta}{\delta i\hat{C}^{ab}}\,e^{-\sum_{i}S[\hat{h}_{i},h_{i};\hat{C},C]},
\end{equation}
and Eqs. \eqref{eq:DFT_S} and \eqref{eq:DFT_W},
the average \eqref{eq:AppB_ss_av_1} can be written as, 
\begin{widetext}
\begin{equation}
   \begin{split}
     \sum_{i=1}^{N}\overline{\bigl\langle S_{i}(t_{a})S_{i}(t_{b})\bigr\rangle_{J}} & =
          \int{\cal D}\hat{C}\,{\cal D}C\,
            e^{-\frac{N}{2}\sum_{ab}i\hat{C}^{ab}C^{ab}}
            \frac{\delta}{\delta i\hat{C}^{ab}}\,e^{-NW[\hat{C},C;0,0]},\\
          & =\int{\cal D}\hat{C}\,{\cal D}C\,
          \left[\frac{\delta}{\delta i\hat{C}^{ab}}+NC^{ab}\right]\,e^{-N{\cal L}[\hat{C},C;0,0]}.
\end{split}
\end{equation}
\end{widetext}
The first terms in the square brackets leads to surface
terms and gives no contribution. Thus 
\begin{equation}
\begin{split}
          \frac{1}{N}\sum_{i=1}^{N}\overline{\bigl\langle S_{i}(t_{a})S_{i}(t_{b})\bigr\rangle_{J}}
             & = \int{\cal D}\hat{C}\,{\cal D}C\,e^{-N{\cal L}[\hat{C},C;0,0]}\,C^{ab}
             \\
             &= \langle C^{ab}\rangle.
\end{split}             
\end{equation}

\section{Averages in the DMFT.}
\label{app:eq_class_pot} 
This Appendix shows how the basic relations used in the main text to express averages
over the solution of the DMFT are obtained.  
These are then used to derive the explicit expression of the potential 
$V(\Delta;\Delta_{0})$ and its derivatives.

Given two generic functions
$\phi(x)$ and $\psi(x)$, and their Fourier representation 
\begin{equation}
\begin{split}
    \phi(x) &=\int_{-\infty}^{+\infty}\frac{dk}{2\pi}\ \tilde{\phi}(k)\ e^{-ikx},
    \\
    \psi(x)
              &=\int_{-\infty}^{+\infty}\frac{dk}{2\pi}\ \tilde{\psi}(k)\ e^{-ikx},
\end{split}              
\end{equation}
then 
\begin{widetext}
\begin{align}
       \langle\phi(h^{a})\,\psi(h^{b})\rangle_{\eta} & =
         \int\frac{dk}{2\pi}\frac{dk'}{2\pi}\ \tilde{\phi}(k)\,\tilde{\psi}(k')\,
              \left\langle e^{-ikh^{a}-ik'h^{b}}\right\rangle _{\eta}
              \nonumber \\
      & =\int\frac{dk}{2\pi}\frac{dk'}{2\pi}\ \tilde{\phi}(k)\,\tilde{\psi}(k')\,
      \exp\Bigl[-\frac{1}{2}(\Delta^{aa}k^{2}+\Delta^{bb}k'^{2})-\Delta^{ab}kk'\Bigr]\nonumber \\
     & =\int\frac{dk}{2\pi}\frac{dk'}{2\pi}\ \tilde{\phi}(k)\,\tilde{\psi}(k')\,
         \exp\Bigl[-\frac{\Delta_{0}}{2}(k^{2}+k'^{2})-\Delta kk'\Bigr].
\label{eq:AppC_pp}
\end{align}
\end{widetext}
In the last line we use $\Delta_{0}=\Delta^{aa}=\Delta^{bb}$ and
$\Delta=\Delta^{ab}$ for $a\not=b$. The integral is well defined
because $|\Delta|\leq\Delta_{0}$. Taking the derivative with respect
to $\Delta^{ab}$ brings down a factor $-kk'$, thus 
\begin{equation}
     \frac{\partial}{\partial\Delta^{ab}}\langle\phi(h^{a})\,\psi(h^{b})\rangle_{\eta}
         =\langle\phi'(h^{a})\,\psi'(h^{b})\rangle_{\eta},
\end{equation}
while the derivative with respect to $\Delta^{aa}$ and $\Delta^{bb}$
gives: 
\begin{equation}
     \frac{\partial}{\partial\Delta^{aa}}\langle\phi(h^{a})\,\psi(h^{b})\rangle_{\eta}
       =\langle\phi''(h^{a})\,\psi(h^{b})\rangle_{\eta},
\label{eq:AppC_der_aa}
\end{equation}
\begin{equation}
      \frac{\partial}{\partial\Delta^{bb}}\langle\phi(h^{a})\,\psi(h^{b})\rangle_{\eta}
          =\langle\phi(h^{a})\,\psi''(h^{b})\rangle_{\eta}.
\label{eq:AppC_der_bb}
\end{equation}
The ``prime'' stands for the derivative of the function with
respect its argument, e.g. $\phi'(x)=(d/dx)\phi(x)$. 

Using the above relations it follows that: 
\begin{equation}
         \begin{split}
          \frac{\partial V(\Delta;\Delta_{0})}{\partial\Delta} 
             & =-\Delta+C(\Delta;\Delta_0)
                          \\
             & =-\Delta+\langle\phi(gh^{a})\phi(gh^{b})\rangle_{\eta}
             \\
             & =\frac{\partial}{\partial\Delta}\left[
                    -\frac{\Delta^{2}}{2}+\frac{1}{g^{2}}\bigl\langle\Phi(gh^{a})\,\Phi(gh^{b})\bigr\rangle_{\eta}
                                                             \right],
\end{split}
\end{equation}
where $\Phi(x)=\int_{0}^{x}dy\,\phi(y)$ is the primitive of the gain
function $\phi(x)$. Integrating  now over $\Delta$ leads to:
\begin{equation}
V(\Delta;\Delta_{0})=-\frac{\Delta^{2}}{2}+\frac{1}{g^{2}}\bigl\langle\Phi(gh^{a})\,\Phi(gh^{b})\bigr\rangle_{\eta}+\mbox{Constant},
\end{equation}
while taking successive derivatives, 
\begin{equation}
\begin{split}
       \frac{\partial^{n}V(\Delta;\Delta_{0})}{\partial\Delta^{n}} =
                    &-\frac{\partial^{n}}{\partial\Delta^{n}}\left(\frac{\Delta^{2}}{2}\right) \\                    
                    &+g^{2n-2}\bigl\langle\phi^{(n-1)}(gh^{a})\,\phi^{(n-1)}(gh^{b})\bigr\rangle_{\eta},
\end{split}
\end{equation}
where $\phi^{(n)}(x)=(d/dx)^{n}\phi(x)$.

The expressions in the main text are obtained by substituting 
\begin{equation}
\begin{split}
    \tilde{\phi}(k)&=\int_{-\infty}^{+\infty}dx\,\phi(x)\ e^{ikx},
    \\
    \tilde{\psi}(k)&=\int_{-\infty}^{+\infty}dx\,\psi(x)\ e^{ikx},
\end{split}    
\end{equation}
into Eq. \eqref{eq:AppC_pp} and performing the resulting Gaussian integrals over the wavenumber:
\begin{align}
      \langle\phi(h^{a})\,&\psi(h^{b})\rangle_{\eta}=\int  \frac{dx\,dy}{2\pi\sqrt{\Delta_{0}^{2}-\Delta^{2}}}\,
              \phi(x)\,\psi(y)\,
      \\
           & \times\exp\Bigl\{-\frac{1}{2(\Delta_{0}^{2}-\Delta^{2})}\bigl[\Delta_{0}(x^{2}+y^{2})
                              -2\Delta\, xy\bigl]\Bigr\}.\nonumber 
\end{align}
Introducing an auxiliary Gaussian variable $z$ the average can be
further written as the integral over independent Gaussian variables:
\begin{align}
      \langle\phi(h^{a})\,\psi(h^{b})\rangle_{\eta} 
      & =\int Dz\int Dx\,\phi(\xi)\int Dy\,\psi(\epsilon_{\Delta}\zeta)\nonumber \\
      & =\int Dz\int Dx\,\phi(\epsilon_{\Delta}\xi)\int Dy\,\psi(\zeta),
\end{align}
where $\epsilon_{\Delta}=\mbox{sign}(\Delta)$, $\xi=\sqrt{\Delta_{0}-|\Delta|}\,x+\sqrt{|\Delta|}\,z$,
$\zeta=\sqrt{\Delta_{0}-|\Delta|}\,y+\sqrt{|\Delta|}\,z$ and $Dz=dz\exp(-z^{2}/2)/\sqrt{2\pi}$
is the Gaussian measure. Notice that if the functions $\phi(x)$ and
$\psi(x)$ have a definite parity then the average vanishes unless
they have the same parity. 
Taking $\psi(x)=\phi(x)$ we recover the expression \eqref{eq:C_DD0} of $C(\Delta,\Delta_{0})$
given in the main text.

If the gain function is odd, as the case discussed in the main text, 
then from the above expressions it easily follows that:
\begin{equation}
       \left.\frac{\partial^{n}}{\partial\Delta^{n}}V(\Delta;\Delta_{0})\right|_{\Delta=0}=0,
       \qquad n=\ {\rm odd}.
\end{equation}
This also implies that $\Phi(x)$ is even, and hence the potential $V(\Delta;\Delta_{0})$ reads: 
\begin{equation}
    V(\Delta;\Delta_{0})=-\frac{\Delta^{2}}{2}+\frac{1}{g^{2}}\int Dz\left[\int Dx\,\Phi(g\xi)\right]^{2}
       +\mbox{Constant}.
\end{equation}

\section{Proof of Eq. \protect\eqref{eq:lyap2}}
\label{app:lyap2}
The four-point correlation in Eq. \eqref{eq:lyap2} can be evaluated following the same
procedure as in \ref{app:eq_C_ab} using the identity:
\begin{equation}
  e^{-S[\hat{C},C, \hat{h}, h]}\sum_{i,j} S_i^{\bs{a}} S_i^{\bs{b}} i\hat{h}_j^{\bs{c}} i\hat{h}_j^{\bs{d}}
         = 
         \frac{\delta}{\delta i\hat{C}^{\bs{ab}}}\, \frac{\delta}{\delta C^{\bs{cd}}}\, 
            e^{-S_[\hat{C},C, \hat{h}, h]}.
\end{equation}
Then 
\begin{equation}
\begin{split}
\frac{1}{N}\sum_{i,j}
  \langle S_{i}^{\bs{a}}S_{i}^{\bs{b}}i\hat{h}_{j}^{\bs{c}}i\hat{h}_{j}^{\bs{d}}\rangle
= & \frac{1}{N}\int  {\cal D}\hat{C}\,{\cal D}C\,
    e^{-\frac{N}{2}\sum_{\mathbf{ab}}i\hat{C}^{\mathbf{ab}}C^{\mathbf{ab}}}
    \\
    & \times
    \frac{\delta}{\delta i\hat{C}^{\mathbf{ab}}}\,\frac{\delta}{\delta C^{\mathbf{cd}}}\,
       e^{-NW[\hat{C},C,0,0]}.
\end{split}       
\end{equation}
Integrating by parts, since the surface terms do not contribute, 
\begin{widetext}
\begin{align}
\frac{1}{N}\sum_{i,j}
  \langle S_{i}^{\bs{a}}S_{i}^{\bs{b}}i\hat{h}_{j}^{\bs{c}}i\hat{h}_{j}^{\bs{d}}\rangle
 & =\int{\cal D}\hat{C}\,{\cal D}C\,C^{\bs{ab}}\,
  e^{-\frac{N}{2}\sum_{\bs{ab}}i\hat{C}^{\bs{ab}}C^{\bs{ab}}}
      \frac{\delta}{\delta C^{\bs{cd}}}\,e^{-NW(\hat{C},C,0,0)}
      \nonumber \\
 & =\int{\cal D}\hat{C}\,{\cal D}C\,
     \Bigl[NC^{\bs{ab}}\,i\hat{C}^{\bs{cd}}-\frac{1}{2}\delta_{\bs{ab},\bs{cd}},\Bigr]\,
         e^{-N{\cal L}[\hat{C},C,0,0]}
         \nonumber \\
 & =N\bigl\langle C^{\bs{ab}}\,i\hat{C}^{\bs{cd}}\bigr\rangle
                     -\frac{1}{2}\delta_{\bs{ab},\bs{cd}},
\end{align}
\end{widetext}
where $\delta_{\bs{ab},\bs{cd}}=\delta_{\bs{ac}}\delta_{\bs{bd}}
        +\delta_{\bs{ad}}\delta_{\bs{bc}}$ is the symmetrised delta and
$\delta_{\bs{ab}} = \delta^{\rm Kr}_{\alpha\beta}\, \delta(t_a - t_b)$.

Alternatively one my notice that
\begin{equation}
   \int{\cal D}\hat{C}\,{\cal D}C\frac{\delta}{\delta i\hat{C}^{\bs{ab}}}\,
    \frac{\delta}{\delta C^{\bs{cd}}}\,e^{-N{\cal L}[\hat{C},C;0,0]}=0
\end{equation}
because this is a surface term and vanishes. On the other hand, from
the form \eqref{eq:DFT_S_r} of ${\cal L}[\hat{C},C;0,0]$ we obtain
\begin{widetext}
\begin{equation}
\begin{split}
         \int{\cal D}\hat{C}\,&{\cal D}C  \frac{\delta}{\delta i\hat{C}^{\bs{ab}}}\,
                                                      \frac{\delta}{\delta C^{\bs{cd}}}\,e^{-N{\cal L}[\hat{C},C;0,0]}
        \nonumber\\                                                      
            & =\int{\cal D}\hat{C}\,{\cal D}C\frac{\delta}{\delta i\hat{C}^{\bs{ab}}}
               \left[\sum_{j}i\hat{h}_{j}^{\bs{c}}i\hat{h}_{j}^{\bs{d}}-Ni\hat{C}^{\bs{ab}}\right]\,
                  e^{-N{\cal L}[\hat{C},C;0,0]}
    \nonumber\\     
       & =\int{\cal D}\hat{C}\,{\cal D}C
               \left[\sum_{j}i\hat{h}_{j}^{\bs{c}}i\hat{h}_{j}^{\bs{d}}-Ni\hat{C}^{\bs{cd}}\right]
               \left[\sum_{j}S{}_{j}^{\bs{a}}S{}_{j}^{\bs{b}}-NC^{\bs{ab}}\right]\,
               e^{-N{\cal L}[\hat{C},C;0,0]}
               \nonumber\\
               &\phantom{===========}
                    -\frac{N}{2}\,\delta_{\bs{ab},\bs{cd}}             
\end{split}                  
\end{equation}
\end{widetext}
Thus, since cross terms vanishes,
\begin{equation}
  \frac{N}{2}\, \delta_{\bs{ab},\bs{cd}}=
      \sum_{ij}\bigl\langle S{}_{j}^{\bs{a}}S{}_{j}^{\bs{b}}i\hat{h}_{j}^{\bs{c}}i\hat{h}_{j}^{\bs{d}}\bigr\rangle
         - N^2 \bigl\langle C^{\bs{ab}}\,i\hat{C}^{\bs{cd}}\bigr\rangle,
\end{equation}
i.e.,
\begin{equation}
      \frac{1}{N}\sum_{i,j}
       \bigl\langle S_{i}^{\bs{a}}S_{i}^{\bs{b}}i\hat{h}_{j}^{\bs{c}}i\hat{h}_{j}^{\bs{d}}\rangle
         =N\bigl\langle C^{\bs{ab}}\,i\hat{C}^{\bs{cd}}\bigr\rangle
           -\frac{1}{2}\delta_{\bs{ab,}\bs{cd}}.
\end{equation}

\newpage{}


\begin{thebibliography}{99}
\bibitem{Note:Causality}
Causality requires that the inverse of the operator $d/dt$ vanishes for $t < 0$
so that the solution of $(d/dt) h = f(h)$ depends on the value of $h(t)$ at previous time.
The term $\delta = 0^+$ ensures that 
$\partial_a^{-1} = e^{-\delta t_a}\theta(t_a)$, where $\theta(t)$ is the Heaviside theta function.
\bibitem{Note:Delta}
Once the time is discretised the (discrete) index $a$ 
denotes the interval $a\delta t \leq t_a <  (a+1)\delta t$.
The Kronecker delta $\delta^{\rm Kr}_{ab}$ is equal to $1$ if the continuous
time variables $t_a$ and $t_b$ fall inside the same time interval of width $\delta t$, and $0$ otherwise.
In the limit $\delta t \to 0$ the width of the interval goes to zero and 
$\delta^{\rm Kr}_{ab} / \delta t \to \delta(t_a-t_b)$, where $\delta(x)$ is the Dirac delta function.
\end{thebibliography}
\end{document}